 \documentclass[final,1p,times]{elsarticle}
\usepackage{graphicx}
\usepackage{amssymb}
\usepackage{amsthm}

\usepackage[nodots]{numcompress}

\usepackage{lineno}

\usepackage{pgfplots}
\usepackage{float}
\newlength\fheight
\newlength\fwidth
\pgfplotsset{compat=newest}

\usepackage{amsmath}

\usepackage{bm}

\usepackage{subcaption}
\usepackage{mwe}

\usepackage[hyphens]{url}
\usepackage{varioref}
\usepackage{hyperref}
\usepackage{cleveref}

\usepackage{dsfont}

\journal{IJEPES}

\begin{document}

\begin{frontmatter}



\title{A rotary frequency converter model for electromechanical transient studies of 16$\frac{2}{3}$~Hz railway systems}

\author[label1]{John Laury\corref{cor1}}
\address[label1]{Lule{\aa} University of Technology, Electrical Power Engineering Group, Skellefte{\aa}, Sweden}
\cortext[cor1]{Corresponding author}
%\fntext[label13]{The financial support for this project from the Swedish Transport Administration is greatly acknowledged.}
\ead{john.laury@ltu.se}
\author[label1]{Lars Abrahamsson}
\author[label1]{Math H. J Bollen}

%\author{John Laury, Lars Abrahamsson and M.H.J Bollen}

\begin{abstract}
%\begin{linenumbers}
Railway power systems operating at a nominal frequency below the frequency of the public grid (50 or 60~Hz) are special in many senses. One is that they exist in a just few countries around the world. However, for these countries such low frequency railways are a critical part of their infrastructure.

%Stability studies published about low frequency railways are few compared to stability studies for the public grid, especially for the 16$\frac{2}{3}$~Hz Scandinavian synchronous ones. One important reason for this is the lack of open models for rotary frequency converters, a component whose behaviour is essential for stability of low-frequency railways.

%The number of published dynamic models as well as stability studies using such -- regarding low frequency railways is small, compared to corresponding publications regarding 50 Hz/60 Hz public grids. This is the case, especially for the 16$\frac{2}{3}$~Hz Scandinavian synchronous ones. One important reason for this is the lack of open models for rotary frequency converters, a component whose behaviour is essential for stability of low-frequency railways.

The number of published dynamic models as well as stability studies regarding low frequency railways is small, compared to corresponding publications regarding 50 Hz/60 Hz public grids. Since there are two main type of low frequency railways; synchronous and asynchronous, it makes the number of available useful publications even smaller. One important reason for this is the small share of such grids on a global scale, resulting in less research and development man hours spent on low frequency grids.

%\textbf{INTRO}: Another important reason is that until railway traffic started to grow again, because of environmental awareness, many people considered the topic of railway electrification as a mature and already developed field.

%Furthermore, transient stability studies are few and there is no open models for rotary frequency converters.

This work presents an open model of a (synchronous-synchronous) rotary frequency converter for electromechanical stability studies in the phasor domain, based on established synchronous machine models. The proposed model is designed such that it can be used with the available data for a rotary frequency converter.

The behaviour of the model is shown through numerical electromechanical transient stability simulations of two example cases, where a fault is cleared, and the subsequent oscillations are shown. The first example is a single-fed catenary section and the second is doubly-fed catenary section.
%two examples of 16$\frac{2}{3}$~Hz railway grids that are synchronous to the 50~Hz public grid.
%Modelling simplifications have been made to simplify the interfacing of rotary frequency converter with railway grid of 16$\frac{2}{3}$~Hz and public grid.

%The paper provides models an insight in the modelling and behaviour of rotary frequency converters supplying railway grids of 16$\frac{2}{3}$~Hz, during and after a severe disturbance, like a fault on the railway side.

%\end{linenumbers}
\end{abstract}


\end{frontmatter}

% \linenumbers

%\begin{linenumbers}
\section{Introduction} \label{Sec: Introduction}
Low-frequency AC railways exist only in six countries: Austria, Germany, Switzerland, Norway, Sweden and in the (North East of the) U.S. \cite{Steimel2012,steimel2008}. As the frequency in the railway is different from the public grid, frequency conversion is needed \cite{steimel2008,Pfeiffer1997}. The conversion can be done by using Motor-Generator sets, also called Rotary Frequency Converter (RFC).

Such an RFC consists of a three-phase motor and a single-phase synchronous generator mounted on the same mechanical shaft.

In Austria, Germany and Switzerland a double-fed induction motor is used, allowing active power to be controlled \cite{steimel2008,Pfeiffer1997,Zynovchenko2005}. The active power supplied by the RFCs follows an active-power-frequency droop characteristic \cite{Zynovchenko2005}.

%and active power can be controlled \cite{steimel2008,Pfeiffer1997}. The frequency of the Austrian, German and Swiss railway grids is controlled to 16.7~Hz.

In Sweden, Norway and the North Eastern U.S. the motor is of synchronous type and therefore the railway grid is synchronous to the three-phase public grid. Active power through an RFC with synchronous motor is dependent on the angle difference between the three-phase public grid and the single-phase low-frequency railway grid of 16$\frac{2}{3}$~Hz (Sweden, Norway) or 25~Hz (North Eastern U.S.) at the RFC locations.

A very limited amount of previous published work has been done on electromechanical stability of low-frequency railways that are synchronous with the public grid.

Small-signal studies on the Norwegian synchronous low-frequency railway grid have been performed in \cite{Danielsen2010a, Danielsen2009, DanielseenTron2009, Danielsen2010}. It was found from those studies that one of the most commonly used RFCs has a poorly damped eigenfrequency that could be excited by modern locomotives, which could lead to system instability. Those studies either use the classical model (which is essentially a constant electromotive force (emf) behind a transient reactance) for each of the two synchronous machines, or the commercial software Simpow for simulations with higher order synchronous machine models.

% -------------------------------------------------
% Fundera till kappan:
% Vilka referenser är klassisk, vilka är simpow?
% -------------------------------------------------

% Exactly what model SIMPOW uses is not clear and available for the public use. The classical model
%Transient stability studies of low-frequency railways published are extremely few. For example,

In \cite{Eitzmann1997} a transient stability assessment is done for the low-frequency railway grid of the North Eastern U.S. No models of RFC were presented in that study as the commercial software PSLF from GE was used.

Reference \cite{Olofsson1989} uses the classical model of a synchronous machine to investigate the transient stability of the Northern part of the Swedish railway system in the end of the 1980's.

There are only a small number of commercial software packages available that simulate dynamics of low-frequency railways that are synchronously connected to the public grid. Furthermore the models used in such software are not published. Thus, complete electromechanical model descriptions of RFCs as part of low-frequency railway grids is non-existent in the literature.

%There are only a few commercial software packages available that simulate dynamics of low-frequency railways that are synchronously connected to the public grid. Furthermore the models used in such software are not published. Thus, complete model descriptions of RFCs as part of low-frequency railway grid is non-existent in the literature.

With increased traffic and with more advanced train control, there is a need for stability studies to increase the understanding of low-frequency railway grids under different scenarios, such as faults or the impact of different control systems. Therefore, there is an identified need for an open dynamic model of an RFC that can be used for transient stability studies. An open, transparent model allows research and education (gathering and spreading of knowledge) on the stability of the mentioned railway without being depended on specialized commercial software packages.
%The commercial software that allows for simulations of low-frequency railway grid synchronous to the public grid are SIMPOW and the 2017 version of PowerFactory.

The aim of this paper is to present a high order synchronous-synchronous RFC model that can be used for transient stability studies. The established 6:th order Andersson-Fouad synchronous machine model \cite{Machowski2008,Milano2010,Fouad2003} is used to describe the dynamics of motor and generator of an RFC.  %The transient stability studies carried out in this paper uses the same methodology as in transient stability studies done in three-phase power systems, but with two grid of different frequencies.
The behaviour of the RFC model is shown in two transient stability studies: a study in which only one RFC is feeding a catenary section (single feeding mode); and a study in which two RFCs are feeding the a catenary section (interconnected mode). The studied results give an insight in the transient stability of synchronously connected low-frequency railway grids.

The remainder of this paper is organised as follows: Section 2 presents both steady state and the proposed dynamical model of an RFC.
Section 3 presents the interfacing of the dynamic RFC machine model to the static grid model for performing the intended electromechanical transient stability studies. The electromechanical transient stability case studies using the proposed RFC model from Section 2 and 3 are presented in Section 4. The results are presented in Section 5 and in Section 6 the conclusions are summarized.

%The models presented are validated through electromechanical transient stability simulations in MatLab Simulink.

%The numerical simulations are done in MATLAB and the behavior of the systems are analyzed.

%The simulations are done to study the behavior of a low-frequency railway with, both in single feeding- and in interconnected mode during and after a severe disturbance.

%The model developed is used in numerical simulations in MatLab, to characterize the behaviour of %one of the most common RFC used in Sweden, both in single feeding- and in interconnected mode %during a severe disturbance.

%There are only a few specialized commercial software packages available that simulate dynamics of low-frequency railways that are synchronously-connected to the public grid. Furthermore, the models used in such software are not open for public use and the software is not always available for research or education, and only limited information if found on RFCs in the literature.
%Furthermore, transient stability for this kind of railway grid is not well explored in the literature.
% to increase the knowledge about this systems.

%% Förslag...
%The RFC model is used in two transient stability case studies to show that the model works as intended, and give an insight of the transien stabily of low-frequency railaywas synchrnous to the public grid. The cases are in single feeding mode and in interconnected mode. The results presented in this paper are analyzed, and give an insight of the transient stability of railway grids that are synchronous to the public grid. 

\section{RFC model}	\label{Sec: Models}

The motors and the generators of the RFCs in the Norwegian and Swedish low-frequency railway grids of 16$\frac{2}{3}$~Hz are all salient pole synchronous machines \cite{Olofsson1989,olofsson1996}. The three-phase motor has $p^m \in 2\cdot\mathds{N}$ magnetic poles, whereas the single-phase generator has $p^{g} = \frac{p^{m}}{3} \in 2\cdot\mathds{N}$ magnetic poles. Thus, it is needed that $p^m \in 6\cdot\mathds{N}$ magnetic poles. The result is that the electrical frequency induced in the stator of the single-phase generator is exactly one third of the frequency in the three-phase public grid of 50~Hz in steady state, that is 16$\frac{2}{3}$~Hz.

Since the models work in per-unit angular frequencies, the exact number of poles in the machines is not of relevance for the computations, but in the Scandinavian system the motors have 12 poles (6 pole pairs), and the generators have 4 poles (2 pole pairs) \cite{olofsson1996}.

%\subsection{Simplifications}

%The step-down transformer reactance is included in the motor reactance and the step up transformer reactance is included in generator reactances.
%
%The RFC are assumed to be connected to infinite buses at the motor side. One result is that the motor voltage is kept to one p.u. and reactive power from the motor is zero. As the motor is connected to infinite bus, the angle $\theta_{50}$ of the infinite bus is kept constant, and assumed equal to zero for simplicity.

%The mechanical shaft where the rotor of motor and generator is mounted on is assumed stiff.

%The machine model used considers the electrical damping. Therefore the damping coefficient $D$ is assumed zero as damping power due windage and mechanical friction is small and therefore neglected.

%Subtransient saliency is neglected, with the assumptions that subtransient reactances are equal in respective machine of the RFC.

\subsection{Steady-state model}
%When a synchronous motor is mechanically loaded, the rotor position will retard relative an idle running synchronous machine. The opposite occurs when a synchronous machine is operated as a generator, the rotor angle will increase in the rotational direction.

%For simplicity, the load angle and rotor angle coincides with each other during steady state, which is the result that leakage reactance is neglected \cite{Saadat2010, Machowski2008}.

From electrical machine theory \cite{Machowski2008,Saadat2010} it is known that the load angle of a salient pole machine
\begin{equation}
\delta = \arctan\left(\frac{X_q P_\blacksquare}{|U|^2 + X_q Q_\blacksquare}\right) \label{Eq 1: Load angle.}
\end{equation}
where $X_q$, $P_\blacksquare$, $|U|$ and $Q_\blacksquare$ are the quadrature reactance, generated/consumed active power, terminal voltage magnitude and generated/consumed reactive power in p.u., respectively. The subscript $\blacksquare$ is used to make \Cref{Eq 1: Load angle.} a generalized expression for motors as well as generators. The load angle defined as in \Cref{Eq 1: Load angle.} is valid under the assumption that the machine resistance is neglected, which is justified by the fact that the resistance is much
smaller than the reactance. In the remainder of this paper, it is for simplicity assumed that the RFCs are lossless.

When an RFC is loaded on the railway grid (generator) side, the total resulting negative (confer \Cref{eq:_sign_of_psi}) voltage phase angle shift between the points of common connection of the motor and the generator is the sum of the load angles of the motor and the generator. The voltage phase shift expressed in $16\frac{2}{3}$~Hz angles
\begin{equation}
\label{Eq 2: Phaseshift}
\psi = \frac{1}{3} \arctan\left(\frac{ - X_{q}^{m}\cdot P_G^{m}}{{|U^{m}|}^2 - X_{q}^{m} Q_G^{m}}\right) + \arctan\left(\frac{X_{q}^{g} \cdot P_{G}^{g}}{{|U^{g}|}^2 + X_{q}^{g} Q_G^{g}}\right),
\end{equation}
where the superscripts $m$ and $g$ stands for motor and generator, respectively.

Positive active power demand on the railway side results in the generated active power of the RFC generator,
\begin{equation}\label{eq:_Positive_generation_in_generator}
P^g_G > 0,
\end{equation}
which means that the generator operates in generator mode. At the same time, the generated active power of the RFC motor
\begin{equation}\label{eq:_Negative_generation_in_motor}
P^m_G < 0,
\end{equation}
which means that the motor operates in motor mode. With the assumption that the RFC is lossless,
\begin{equation}
\label{eq:_Relation_generation_motor_generator}
-P^m_G = P^g_G
\end{equation}
holds.

For the simplicity of modelling, the step-down transformer leakage reactance of the motor $X_T^m$ (confer \Cref{Tab 1: Motor Parameters}) is added to $X^m_q$ in order to redefine $X^m_q$ according to,
\begin{equation}
\label{eq:_Reactances_Added}
X^m_q    := X_T^m + X_q^m,
\end{equation}
and the step-up transformer leakage reactance $X_T^g$ of the generator is added to $X^g_q$ in order to form a new redefined $X^g_q$ in the same way as \Cref{eq:_Reactances_Added} for the motor.
%Note that negative power flow is into the motor and positive power flow is out from the generator.
%, which results that $Pg_M\leq 0$.

%Note that $Pg_M\leq 0$ as the motor acts as an active power consumer for the three-phase grid. The division with 3 in \Cref{Eq 2: Phaseshift} results that voltage phase shift is expressed in $16\frac{2}{3}$Hz angle.

As active power flows from the public grid to the railway grid, the terminal angle of the single-phase generator will fall relative to the angle $\theta^m_{50}$ on the three-phase grid to which the motor is connected. This angle drop takes place in two steps; first
\begin{equation}
\label{eq:_angle_shift_step1}
\delta_m = \frac{\theta_{50}^m}{3} - \frac{1}{3} \arctan\left(\frac{ - X_{q}^{m}\cdot P_G^{m}}{{|U^{m}|}^2 - X_{q}^{m} Q_G^{m}}\right),
\end{equation}
and secondly
\begin{equation}
\label{eq:_angle_shift_step2}
\theta^g = \delta_m - \arctan\left(\frac{X_{q}^{g} \cdot P_{G}^{g}}{{|U^{g}|}^2 + X_{q}^{g} Q_G^{g}}\right),
\end{equation}
where $\delta_m$ denotes the per-unit mechanical angle of \Cref{Eq 15: TM matrix}. The angle $\delta_m$ also represents the electrical angle of the RFC's rotor expressed in the \text{16$\frac{2}{3}$ Hz} grid frame. Therefore, putting \Cref{eq:_angle_shift_step1,eq:_angle_shift_step2,Eq 2: Phaseshift} together, the terminal angle of the generator
\begin{equation}
\label{eq:_sign_of_psi}
\theta^g = \frac{\theta^m_{50}}{3} - \psi.
\end{equation}

At the railway grid side, the voltage magnitude $|U^g|$ after the step-up transformer is controlled as
\begin{equation}
|U^g| = U_0 - K_U Q^g_G,
\label{Eq 3: Droop control}
\end{equation}
where $U_0$ is the no-load voltage. In addition, $K_U$ in \Cref{Eq 3: Droop control} is a droop coefficient which is scaled to the RFC single-phase generator rating. The use of such a scaled droop coefficient $K_U$ results in reactive load sharing between the different active RFCs in a converter station \cite{Laury2017}.

On the three-phase side; there are according to \cite{KravspecOmformareOmriktareTDOK2013_0670} three control options:
\begin{enumerate}
	\item the terminal voltage at the bus of common connection is controlled to a constant value,
	\item the reactive power of the motor is controlled to a constant value, or
	\item the power factor of the motor is controlled to a constant value,i
\end{enumerate}
that can be used. Since the models proposed and applied in this paper assumes an infinite bus connecting to the motors, the resulting control in steady-state will in practice be both options 1 and 2 in the list above, with voltage $U^m$ controlled to 1 \text{p.u.} and reactive power $Q^m_{G}$ controlled to 0 \text{p.u.} Dynamically, however, none such control takes place, since it is not needed. During the transient, the reactive power of the motor will fluctuate a bit until it stabilizes at 0 again.

\subsection{Dynamic model}

\subsubsection{The mechanical equations}

%As was mentioned before, the synchronous machines are mounted on the same mechanical shaft.
The mechanical shaft to which one RFC's both synchronous machines are mounted on is assumed to be mechanically stiff. That assumption implies that the rotors of both machines in one RFC rotate at the same speed. The load torque of the synchronous motor is the input to the synchronous generator. This results in the swing equation of an RFC to be
%The result is that $\omega^M_m = \omega^G_m$ and one state equation is needed for the speed of the rotor. %For simplicity, the indices $M$ and $G$ is dropped for the mechanical angle and speed. thus combining \Cref{Eq 4: Swing equation motor,Eq 5: Swing equation generator} results in
\begin{equation}
(J^m + J^g) \frac{d\omega_m(t)}{dt} = T^m(t) - T^g(t), \label{Eq 6: Swing equation of RFC}
\end{equation}
where $J^m$ and $J^g$ denote the moments of inertia of motor and generator, respectively. Moreover, $\omega_m$ stands for the mechanical angular frequency, whereas $T^m$ and $T^g$ stand for the electromagnetic air gap torques of motor and generator, respectively.

The electrical part of the machine model proposed in this paper and presented in \Cref{subsub Electrical}, includes the effects of damper windings. Mechanical damping caused by windage and friction is assumed to be small and is therefore neglected.

%From \Cref{Eq 6: Swing equation of RFC} it is stated that the moment of inertia of motor and generator are added, and
A synchronously rotating reference frame with a constant mechanical angular frequency, $\omega_{sm}$, is chosen as angular reference, so that the mechanical rotor angle (that is, the angular displacement of the rotor with respect to a stationary axis on the stator),
\begin{equation}
\beta_m(t) = \omega_{sm}t + \delta_m(t),
\label{Eq 7: Relative anlge}
\end{equation}
where $\delta_m\left(t\right)$ is the angular position with respect to the chosen synchronously rotating reference frame $\omega_{sm}$. The velocity (that is, the angular frequency) of the rotor (confer to \Cref{Eq 6: Swing equation of RFC}),
\begin{equation}
\omega_m\left(t\right) = \frac{d\beta_m(t)}{dt}= \omega_{sm}+\frac{d\delta_m(t)}{dt} \label{Eq 8: Angle of delta_m},
\end{equation}
relative to the synchronously rotating reference frame. In addition, the acceleration of the mechanical rotor angle
\begin{equation}
\frac{d^2\beta_m(t)}{dt^2}=\frac{d^2\delta_m(t)}{dt^2}.
\end{equation}
The total moment of inertia of the RFC,
\begin{equation}
J^{mg} = J^m + J^g
\end{equation}
and $J^{mg}$ as expressed in terms of the per unit inertia constant $H^{mg}$ becomes
\begin{equation}
J^{mg} = \frac{2H^{mg}S_B}{\omega_{sm}^2}.
\label{Eq: Intertia constant}
\end{equation}
In \Cref{Eq: Intertia constant}, $S_B$ denotes the base power used and $S_B$ is numerically defined in \Cref{Tab 3: System Parameters}. Inserting \Cref{Eq: Intertia constant} in \Cref{Eq 6: Swing equation of RFC} and multiplying with the mechanical angular speed yields
\begin{equation}
\omega_m(t) \frac{2H^{mg}S_B}{\omega_{sm}^2} \frac{d\omega_m(t)}{dt} = \omega_m(t) \left(T^m(t) - T^g\left(t\right)\right). \label{Eq: No per unit swing}
\end{equation}
Reformulating \Cref{Eq: No per unit swing} and expressing it in \text{p.u.} results in
\begin{equation}
2H^{mg}\omega_{\text{p.u.}}(t) \frac{d\omega_{\text{p.u.}}(t)}{dt} = - P^{m}_{\text{p.u.}}(t) - P^{g}_{\text{p.u.}}(t). \label{Eq 9: p.u. swing}
\end{equation}
Moreover, \Cref{Eq 8: Angle of delta_m,Eq 9: p.u. swing} can be transformed to the state space form,
\begin{align}
\frac{d\delta_m(t)}{dt} &= \Delta \omega_{\text{p.u.}}(t) \omega_{sm} \label{Eq 10: mechanical angle}\\
\frac{d\omega(t)_{\text{p.u.}}}{dt} &= \frac{1}{2H^{mg}\omega(t)_{\text{p.u.}}} (-P^{m}_{\text{p.u.}}(t) - P^{g}_{\text{p.u.}}(t)), \label{Eq 11: deviation}
\end{align}
where
\begin{equation}
\Delta \omega_{\text{p.u.}}(t) = \omega_{\text{p.u.}}(t) - 1
\end{equation}
is the speed deviation of the rotor from the synchronously rotating reference frame, $\omega_{sm}$.

Note that the torque of a motor is defined positive for consumption, whereas the torque of a generator is defined positive for generation. Since this paper consequently treats all machine powers as generated powers, the positive motor torque in \Cref{Eq: No per unit swing} multiplied by the mechanical angular frequency $\omega_m\left(t\right)$ becomes a negatively signed power, that is $ - P^{m}_{\text{p.u.}}(t)$, in \Cref{Eq 9: p.u. swing,Eq 11: deviation}. The air gap powers in \Cref{Eq 9: p.u. swing,Eq 11: deviation} generated in the motor
\begin{equation}\label{eq:_Lossless_Motor}
P^{m}_{\text{p.u.}}(t) = P^m_G\left(t\right),
\end{equation}
and the generator
\begin{equation}
\label{eq:_Lossless_Generator}
P^{g}_{\text{p.u.}}(t) = P^g_G\left(t\right),
\end{equation}
respectively, because of the assumption of lossless machines.
%Note the minus sign in the air gap power $P^{m}_{p.u.}(t)$ , which is because negative power flow is into the motor in motor mode.
The mechanical rotor angle (to be stringent; the angular position with respect to $\omega_{sm}$), $\delta_m$, is multiplied by the number of pole pairs $\frac{p^g}{2}$ and $\frac{p^m}{2}$ to obtain the electrical rotor angle in the $16\frac{2}{3}$~Hz and the 50~Hz electrical frames, respectively.

%where $\Delta \omega_{p.u.}(t) = \omega_{p.u.}(t) - 1 $ is the speed deviation of the rotor from the synchronously rotating reference frame. As negative power flow is into the RFC motor and positive out from the RFC generator, a minus sign is inserted in front of the air gap power $P^{m}_{\text{p.u.}}(t)$ in \Cref{Eq 11: deviation}.

%The mechanical rotor angle $\delta_m$ is multiplied by the number of pole pairs $\frac{p^g}{2}$ and $\frac{p^m}{2}$ to obtain the electrical rotor angle in the $16\frac{2}{3}$~Hz and the 50~Hz electrical frame, respectively.

\subsubsection{The electrical equations} \label{subsub Electrical}

The 6:th order Andersson-Fouad model \cite{Machowski2008,Milano2010,Fouad2003} of a synchronous machine is used to describe the links between stator fluxes, stator currents and field voltages. Such a machine model considers the dynamics of the damper windings and is suitable to be used for stability studies \cite{Fouad2003}. For simplicity magnetic saturation is neglected.

The time derivative of the transient emf in the direct axis
\begin{equation}
\dot{E}'_{d}= -E'_d -I_q(X_q-X'_q),
\label{eq:_EdPrimeDot}
\end{equation}
however, the synchronous machines are of salient pole type, which implies that
\begin{equation}
X'_q = X_q
\label{eq:_QuadReactSaliencyProperty}
\end{equation}
according to \cite[Table 4.2]{Machowski2008}. Since $\dot{E}'_{d}$ is equal to zero in steady-state, \Cref{eq:_EdPrimeDot,eq:_QuadReactSaliencyProperty} give that $E'_d = 0$ for motor as well as generator in a Scandinavian RFC. In turn, the 6:th order model of \cite[Section 11.1.7.1]{Machowski2008} collapses to the 5:th order model of \cite[Section 11.1.7.2]{Machowski2008}.
%for respective machine \cite{Machowski2008,Padiyar2008}.
Note, that all variables henceforth are expressed in the \text{p.u.} system, unless otherwise stated. Therefore, the \text{p.u.} subscripts are omitted henceforth.

The resulting electrical equations of an RFC are:
\begin{align}
T'^{m}_{do}\dot{E}'^{m}_{q} &= E^m_{f} - E'^m_q + I^m_{d}(X^m_{d}-X'^m_{d}) \label{Eq 12: Sync machine eq 1}\\
T''^{m}_{do}\dot{E}''^{m}_{q} &= E'^m_{q} - E''^m_q + I^m_{d}(X'^m_{d}-X''^m_{d}) \label{Eq 13: Sync machine eq 2}\\
\begin{split}
T''^{m}_{qo}\dot{E}''^{m}_{d} & = -E''^m_{d} - I^m_{q}(X'^m_{q}-X''^m_{q}) = \\
& = \left\{ \text{\Cref{eq:_QuadReactSaliencyProperty}} \right\} = \\
& = -E''^m_{d} - I^m_{q}(X^m_{q}-X''^m_{q})
\end{split}
\label{Eq 14: Sync machine eq 3}\\
T'^{g}_{do}\dot{E}'^{g}_{q} &= E^g_{f} - E'^g_q + I^g_{d}(X^g_{d}-X'^g_{d}) \label{EqG 12: Sync machine eq 1}\\
T''^{g}_{do}\dot{E}''^{g}_{q} &= E'^g_{q} - E''^g_q + I^g_{d}(X'^g_{d}-X''^g_{d}) \label{EqG 13: Sync machine eq 2}\\
\begin{split}
T''^{g}_{qo}\dot{E}''^{g}_{d}   & = -E''^g_{d} - I^g_{q}(X'^g_{q}-X''^g_{q}) = \\
& = \left\{ \text{\Cref{eq:_QuadReactSaliencyProperty}} \right\} = \\
& = -E''^g_{d} - I^g_{q}(X^g_{q}-X''^g_{q})
\end{split}
\label{EqG 14: Sync machine eq 3}
\end{align}
where $T'_{do}$ is the direct-axis transient open circuit time constant. The sub-transient open circuit time constant in the quadrature and the direct axes are  $T''_{qo}$ and $T''_{do}$, respectively. The transient voltage in the quadrature axis is $E'_q$, whereas the sub-transient voltage in the direct and the quadrature axes are $E''_d$ and $E''_q$. The steady-state, transient and sub-transient reactances in the direct-axis are $X_d$, $X'_d$ and $X''_d$, respectively. The steady-state, transient and sub-transient reactances in the quadrature-axis are $X_q$, $X'_q$ and $X''_q$, respectively, but since salient pole machines are used \Cref{eq:_QuadReactSaliencyProperty} still holds.

%where $T'_{do}$ is the direct-axis transient open circuit time constant. $T''_{do}$ and $T''_{qo}$ are the direct- and quadrature-axis sub-transient open circuit time constant. $E'_q$ is the transient quadrature-axis voltage; $E''_d$ and $E''_q$ are the sub-transient voltage in direct- and quadrature-axis, respectively. $X_d$, $X'_d$ and $X''_d$ are the steady-state, transient and sub-transient reactance in direct-axis, respectively. Whereas $X_q$, $X'_q$ and $X''_q$ are the steady-state, transient and sub-transient reactance in quadrature-axis, respectively.

The air-gap powers of both synchronous machines of an RFC are:
\begin{align}
P^m &= E''^m_d I^m_{d} + E''^m_{q}I^m_{q} + (X''^m_{d} - X''^m_{q})I^m_{d}I^m_{q} \label{Eq 15: Airgap machine Motor} \\
P^g &= E''^g_d I^g_{d} + E''^g_{q}I^g_{q} + (X''^g_{d} - X''^g_{q})I^g_{d}I^g_{q}. \label{EqG 15: Airgap machine Gen}
\end{align}
%As the RFC consist of two machines, \Cref{Eq 12: Sync machine eq 1,Eq 13: Sync machine eq 2,Eq 14: Sync machine eq 3,Eq 15: Airgap machine} is used the describe the electrical parts for motor and generator respectively.

\subsubsection{Dynamic RFC model}
With \Cref{Eq 10: mechanical angle,Eq 11: deviation,Eq 12: Sync machine eq 1,Eq 13: Sync machine eq 2,Eq 14: Sync machine eq 3,EqG 12: Sync machine eq 1,EqG 13: Sync machine eq 2,EqG 14: Sync machine eq 3,Eq 15: Airgap machine Motor,EqG 15: Airgap machine Gen}, a total of eight first order differential equations and two algebraic equations are obtained to describe the dynamics of an RFC excluding the excitation system model. There are 8 differential equations since 2 machines with 5:th order models are considered, but they share the mechanical properties in \Cref{Eq 10: mechanical angle,Eq 11: deviation} (5+5-2 = 8).

\subsubsection{Excitation system}
To control the voltage magnitude after the step-up transformer on the railway side, the RFC generator is equipped with an excitation system. The inputs are the calculated voltage magnitude reference  $|U|^\text{ref}$ and the measured voltage magnitude $|U^g|$. Deviation from the calculated voltage reference value causes that the excitation system to adjust the field voltage $E^g_f$ of the RFC generator.

On the motor side, the voltage magnitude or reactive power is controlled to a constant reference value. If there is a deviation from the reference value, the excitation system of the motor will adjust the field voltage $E^m_f$.

For modelling purpose, any of the standard models of excitation systems from \cite{IEEE_AVR_2005} can be used on the RFCs motor and generator, respectively.

%Both the motor and generator field voltage can be controlled. This adds additional differential equations to the RFC dynamics, depending of the type of excitation system used.
%The excitation system is on generetor is to controll the voltage magntide of the
%
%As RFCs generator have brushless excitation systems in Sweden. Therfore,  the AC5A \cite{IEEE_AVR_2005} excitation system model from MatLab has been used on the generator side to control the voltage after the step-up transformer according to \ref{Eq 3: Droop control}.

\section{Model for electromechanical transient stability studies}	\label{Sec: Model2}
The stator voltages of each machine are expressed in their own dq rotating reference frame for that machine. This reference rotates independently from other machines of the system.

The electrical equations of the three phase public grid are expressed in a common 50~Hz Re-Im reference frame. The subscript ${a_{50}}$ and ${b_{50}}$ stands for real and imaginary part of that reference frame, respectively. The electrical equations of the railway grid are also expressed in common Re-Im reference frame, but rotating at $16\frac{2}{3}$~Hz. The subscript $a_{16\frac{2}{3}}$ and $b_{16\frac{2}{3}}$ stands for the real and imaginary part of that frame.

The transformation matrix $\bm{T}$ is used to transform the motor and generator stator quantities from a 50~Hz or $16\frac{2}{3}$~Hz reference frame to the rotor reference frame of the particular RFCs motor and generator, respectively. The transformation matrix,
\begin{equation}
\bm{T} =	\begin{bmatrix}
-\sin(\frac{p}{2} \delta_m) & \cos(\frac{p}{2}\delta_m) \\
\cos(\frac{p}{2}\delta_m) & \sin(\frac{p}{2}\delta_m)
\end{bmatrix} \label{Eq 15: TM matrix}
\end{equation}
where $p$ is the number of poles of the particular machine considered. Note that $\bm{T}$ is orthogonal and the same matrix is used for backward and forward transformations.

If $p=p_g$ in \Cref{Eq 15: TM matrix} the transformation matrix $\bm{T}$ is denoted as $\bm{T^g}$, and used to transform the dq rotor variables of the generator to the $16\frac{2}{3}$~Hz frame of the railway grid and the other way around. The transformation matrix $\bm{T}$ is denoted as $\bm{T^m}$ if $p=p_m$, and used to transform the dq rotor variables of the motor to the 50~Hz reference frame.

\begin{figure}
	\centering
	\includegraphics[scale=0.6, keepaspectratio]{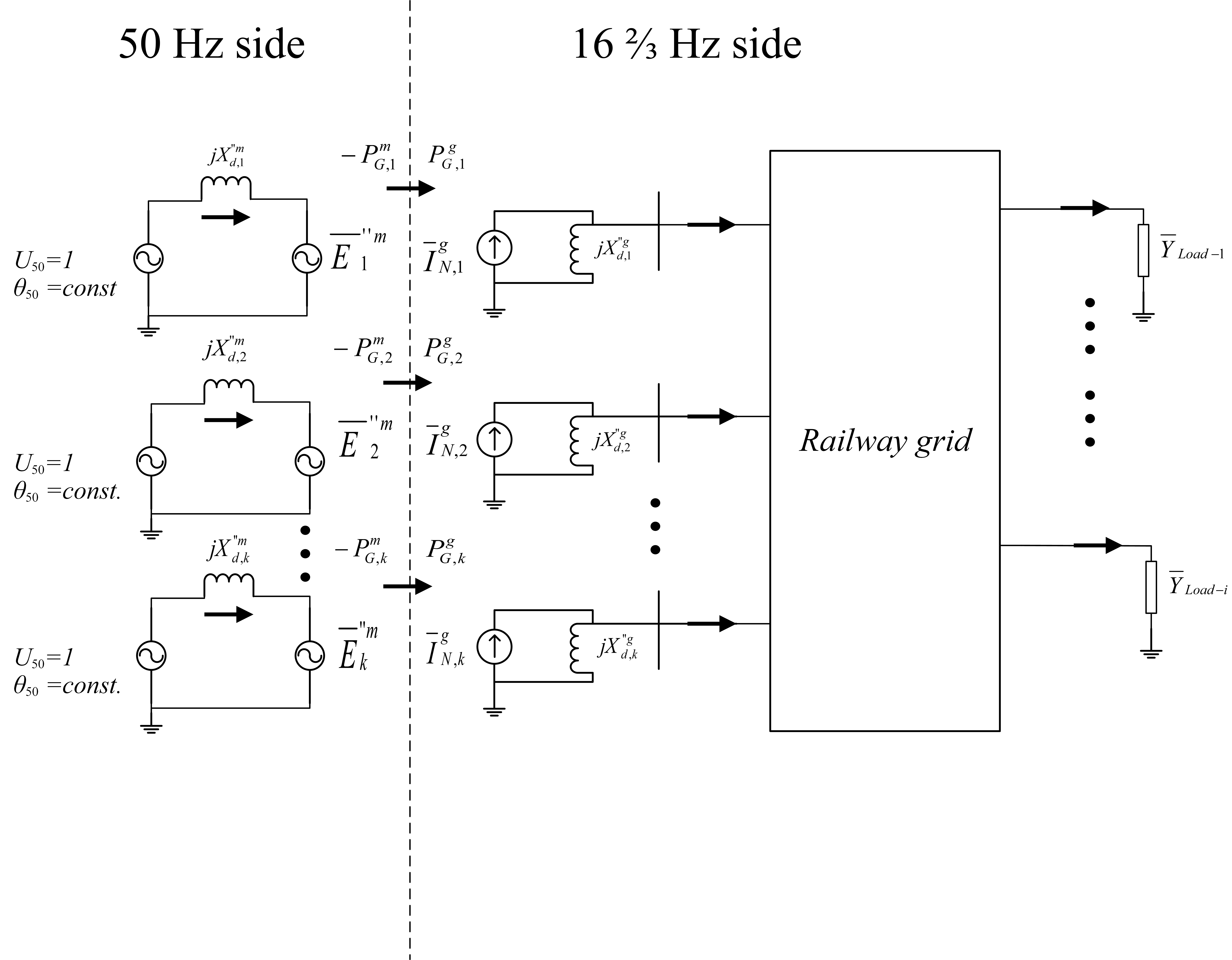}
	\caption{Model setup of a synchronous-synchronous railway system for i= 1...k RFCs.}
	\label{Fig 1: System}
\end{figure}

Consider a synchronous railway system as shown in \Cref{Fig 1: System}. All RFCs motor are connected to infinite buses, where the voltage $U_{50}$ is set to one p.u. and the voltage phase angle $\theta_{50}$ is set zero and is constant. This assumption is justified as several connection points in the public grid where the RFC are connected to are strong. The $i$:th RFC motor is then implemented as a voltage source $\overline{E}''^{m}_{i}=E''^{m}_{d,i} + jE''^{m}_{q,i}$ in the dq rotor reference frame of the motor.

On the railway side, the $i$:th RFC generator is connected to the railway grid as an equivalent Norton current source, $\bar{I}^g_{N,i}$. The voltages $E''^g_{d,i}$ and $E''^g_{q,i}$ of the $i$:th RFC generator are transformed from the dq reference of generator into the $16\frac{2}{3}$~Hz reference frame, and in that frame the injected Norton current is calculated according to
\begin{align}
\begin{bmatrix}
E''^g_{{a,i}_{16\frac{2}{3}}}\\
E''^g_{{b,i}_{16\frac{2}{3}}}
\end{bmatrix}
&=
\bm{T^g_{i}}
\begin{bmatrix}
E''^{g}_{d,i}\\
E''^{g}_{q,i}
\end{bmatrix} \label{Eq 16a: Norton currents} \\
\overline{I}^g_{N,i} &= \frac{E''^g_{{a,i}_{16\frac{2}{3}}}\ + jE''^g_{{b,i}_{16\frac{2}{3}}}}{jX''^g_{d,i}} \label{Eq: 16b}
\end{align}

According to \cite{Machowski2008,Dandeno1973}, the effects of sub-transient saliency are normally small and can therefore be neglected, with the assumption that $X''_d \approx X''_q$. As seen in \ref{Tab 1: Motor Parameters} the approximation is acceptable for the single phase-generator, whereas for the three-phase motor the approximation is not acceptable as there is to large difference between $X''_d$ and $X''_q$. However, to avoid the need of iterative computation for handling saliency and for fast simulations, $X''_d$ is set equal to $X''_q$ for both machines. This assumption affects \Cref{Eq 15: Airgap machine Motor,EqG 15: Airgap machine Gen}. %Setting $X''_d = X''_q$ has also the benefit that the interfacing of the RFCs synchronous machines to public grid and railway grid is simplified for implementation in numerical software.

%Through this assumption, the interfacing  of the RFC synchronous machines to the public grid and railway grid is simplified for implementation in numerical software.
% In the simulation done, the larger of  $X''_d$ and $X''_q$ is used to counterbalance the effect on neglecting sub-transient saliency.

\begin{figure}[h!]
	%\centering
	\includegraphics[width=\textwidth,height=1.5\textheight,keepaspectratio]{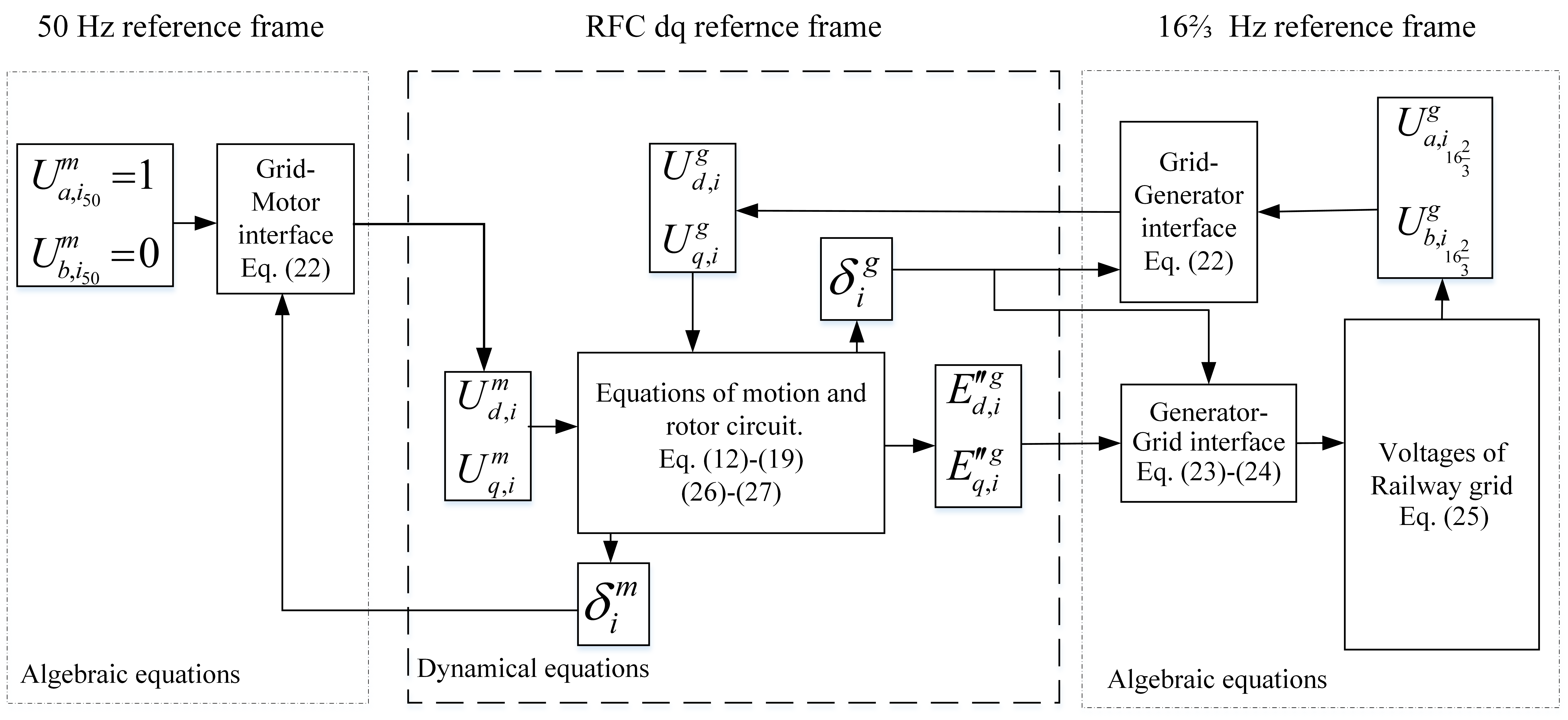}
	\caption{Model structure of the $i$:th RFC connected to both public grid and railway grid.} \label{Fig: flowchart}
\end{figure}

To connect the RFCs' generators to the railway grid for the simulations, the admittance matrix $\bm{Y_{bus,{16\frac{2}{3}}}}$ describing the railway grid is augmented to $\bm{Y^{aug}_{bus,{16\frac{2}{3}}}}$.
The augmentation is done by adding the sub-transient reactances as shunt admittances to the $\bm{Y_{bus,{16\frac{2}{3}}}}$  matrix where the RFC generators are connected to the railway grid. For simplicity, the loads are modelled as shunt admittances and are added to in a similar way to the $\bm{Y_{bus,{16\frac{2}{3}}}}$ matrix.
As $X''_d = X''_q$ for both of the RFC machines and loads are expressed as admittances, a linear load flow is used. The voltages of the railway grid are calculated in the $16\frac{2}{3}$~Hz common Re-Im phasor frame as:
\begin{equation}
\bm{U_{16\frac{2}{3}} = \bm({Y^{aug}_{bus,{16\frac{2}{3}}}})^{-1}}\bm{I^g_N}	\label{Eq 18: Voltages}.
\end{equation}
Solving \Cref{Eq 10: mechanical angle,Eq 11: deviation,Eq 12: Sync machine eq 1,Eq 13: Sync machine eq 2,Eq 14: Sync machine eq 3,EqG 12: Sync machine eq 1,EqG 13: Sync machine eq 2,EqG 14: Sync machine eq 3,Eq 15: Airgap machine Motor,EqG 15: Airgap machine Gen,Eq 18: Voltages}, the current of the $i$:th RFCs generator and motor in the dq frame, respectively is
\begin{align}
\begin{bmatrix}
I^g_{d,i}\\
I^g_{q,i}
\end{bmatrix}
&=
\frac{1}{X''^g_{d,i}}
\begin{bmatrix}
U^g_{q,i} - E''^g_{q,i}\\
-U^g_{d,i} + E''^g_{d,i}
\end{bmatrix}\\
\begin{bmatrix}
I^m_{d,i}\\
I^m_{q,i}
\end{bmatrix}
&=
\frac{1}{X''^m_{d,i}}
\begin{bmatrix}
U^m_{q,i} - E''^m_{q,i}\\
-U^m_{d,i} + E''^m_{d,i}
\end{bmatrix}.
\end{align}
Active and reactive power injected by the $i$:th generator of the RFC into the railway grid is computed as:
\begin{align}
P^g_{i} &= U^g_{d,i}I^g_{d,i} +  U^g_{q,i}I^g_{q,i} \label{Eq 20: RFCgen Active Power} \\
Q^g_{i} &= U^g_{d,i}I^g_{q,i} -  U^g_{q,i}I^g_{d,i} \label{Eq 21: RFCgen Reactive Power}
\end{align}
Active and reactive power of the motor injected by the $i$:th motor of the RFC into the 50 Hz grid is computed as:
\begin{align}
P^m_{i} &= U^m_{d,i}I^m_{d,i} +  U^m_{q,i}I^m_{q,i} \label{Eq 22: RFCgen Active Power} \\
Q^m_{i} &= U^m_{d,i}I^m_{q,i} -  U^m_{q,i}I^m_{d,i}. \label{Eq 23: RFCgen Reactive Power}
\end{align}
The computational structure of $i$:th RFC connected to both the railway grid and public grid is shown in \Cref{Fig: flowchart}.

The electrical rotor angles of the motors are referred to the constant 50 Hz angle $\theta_{50}$ of the infinite bus. The electrical rotor angles of the generators in interconnected mode can be referred to any of the other RFC generator on the railway grid side or the angle of the 50~Hz side.

%or to the infinite bus on the 50~Hz side where the motor is connected to.

\subsection{Modelling discussion}

In the overall modelling, transients associated with power/current flows in the railway grid and the public grid are assumed to have a fast decay time, so that a QSS (Quasi Steady-State) \cite{CutsemVournasRedBook,AbrahamssonRiskFullReport} approach is used. As grid transients are neglected, the stator transients of the RFC machines have to be neglected \cite{Kundur1994,Machowski2008}.

%In an example illustrating the impact of the negligence introduced in the above paragraph for a public grid system (that is, a grid fed by a generator in the \textit{normal} way by a governor controlling the mechanical power applied to the shaft of it) can be found in \cite[Section 5.1.1]{Kundur1994}. In that example, it is shown that the negligence of stator transients makes the transients of $I_d$, $I_q$, and $T$ non-oscillatory, whereas they oscillate after a cleared fault without this negligence. Moreover, in that example, the rotor angle oscillations are shown to be exaggerated and slightly slower making this negligence. Also, the rotor speed deviation becomes greater using this negligence than not doing so; and the oscillatory behaviour of the rotor speed deviation before fault clearing vanishes using this negligence, according to the example in \cite[Section 5.1.1]{Kundur1994}.

%For public power systems,

The negligence of stator transient introduces a conservatism to the model, as rotor speed deviation is increased \cite{Kundur1994}. One should however proceed with care regarding the implications of this negligence for RFC-fed synchronously operated (single-phase) low frequency railway grids, since so many operating conditions differ from public transmission grids fed by thermal or hydro-power generators. For example, the mechanic power of the generator shaft is provided by another synchronous machine for synchronous-synchronous railway systems.

%as can be seen in both \cite{LauryJRC18} and in \Cref{Fig 6: Case 1 rotor speeds,Fig 9: Case 2 Rotor speeds}, the characteristics of rotor speed deviations in RFC-fed synchronous low frequency railway grids look totally different than the ones of public power systems. One important reason for this is that

As recommended in \cite{Machowski2008,Kundur1994}, also the rotor speed deviations are neglected in the stator equations. According to \cite{Kundur1994}, this is not mainly done to make computations faster, but rather to counterbalance the effect of stator transients being neglected, considering the low-frequency rotor oscillations.

% The assumption of nominal rotor speed in the stator equations are according to \cite[Section 5.1.2]{Kundur1994} valid as long as the speed deviations are small and have no significant effect on the voltage levels. Small deviations in rotor speed are also mentioned as an argument in \cite[Section 11.1.4]{Machowski2008}. In \cite{LauryJRC18} one can see that rotor speeds never deviate more than 2\%, and in \Cref{Fig 6: Case 1 rotor speeds,Fig 9: Case 2 Rotor speeds}, they never deviate more than 3\%. In some cases in \cite{LauryJRC18}, like in \cite[Figure 31]{LauryJRC18} one can see a slightly larger, but still small, impact of rotor oscillations on voltage levels.

\section{Cases studies} \label{Sec: Simulations}

\begin{figure}
	\centering
	\includegraphics[scale=0.65]{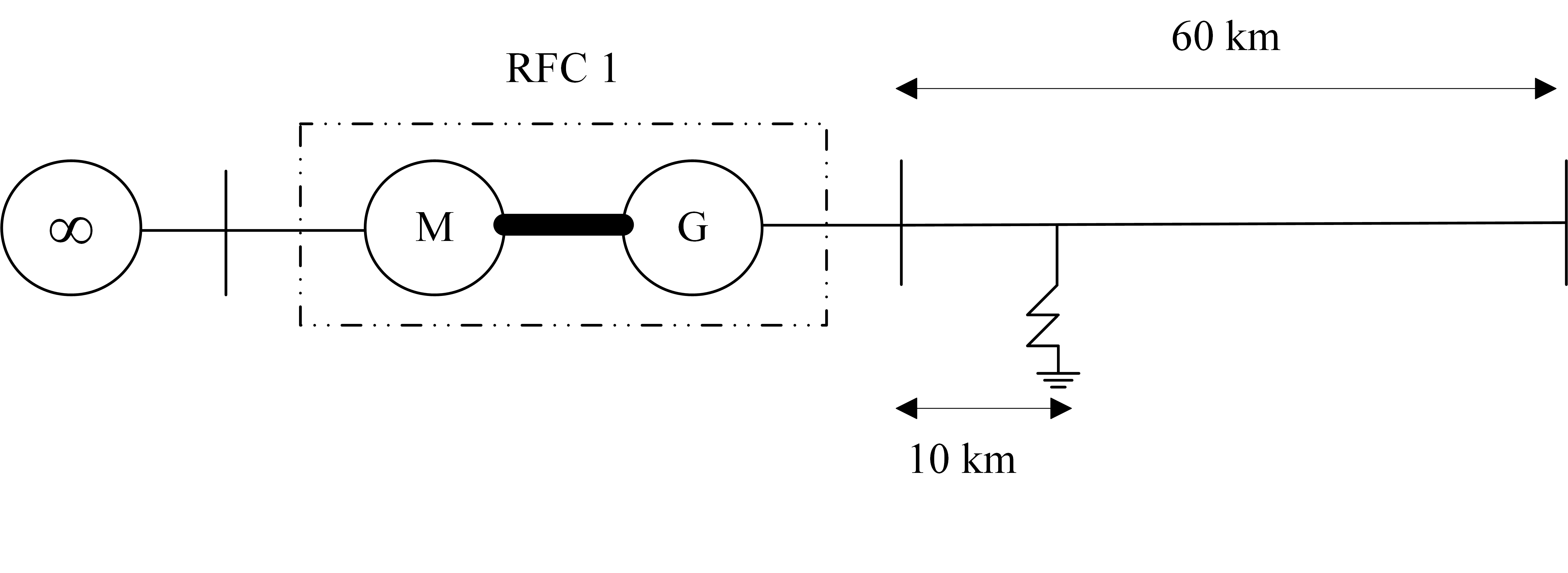}
	\caption{Case 1: Single feeding mode.}
	\label{Fig 2: System one RFC}
\end{figure}

\begin{figure}
	\centering
	\includegraphics[scale=0.65]{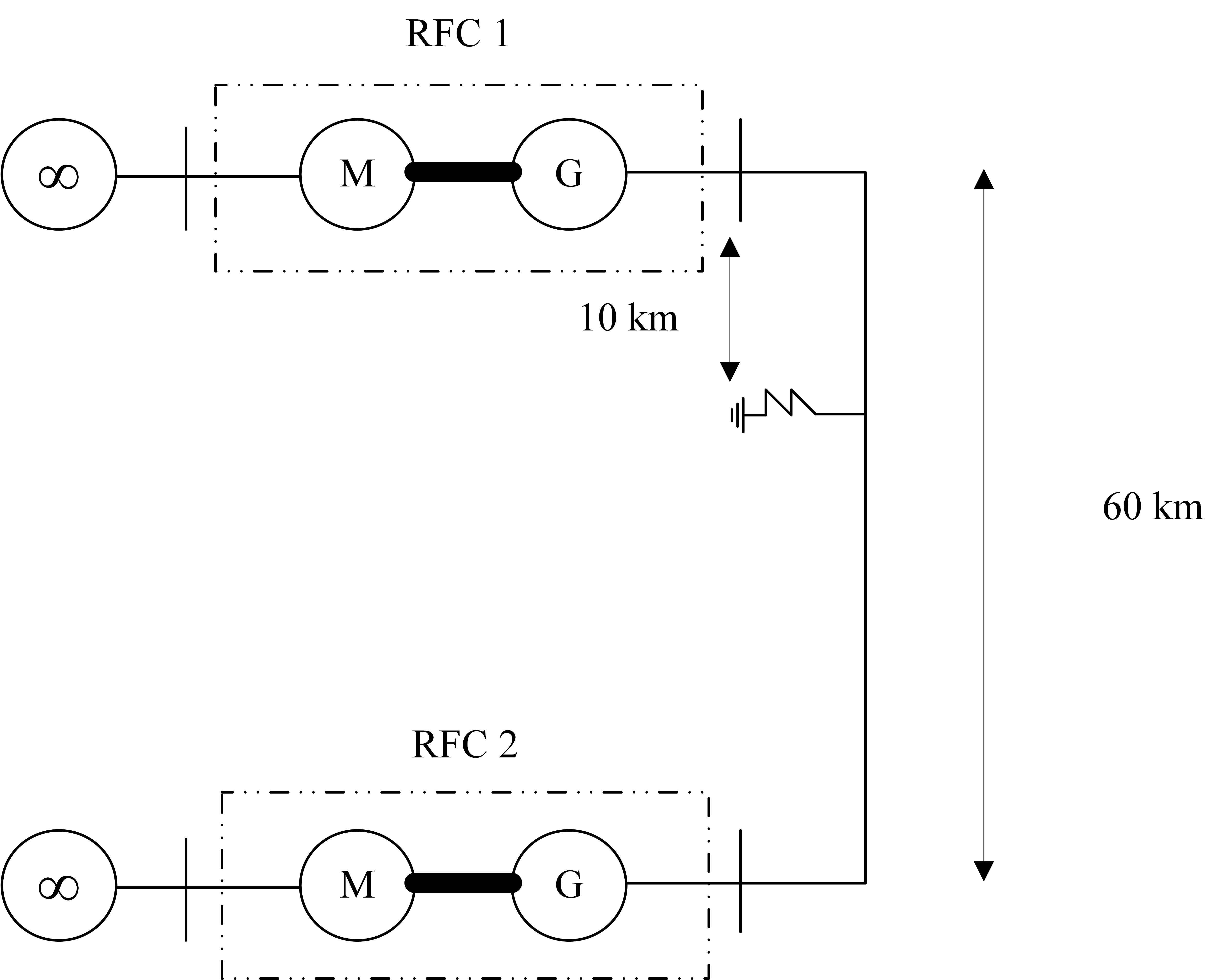}
	\caption{Case 2: Interconnected mode.}
	\label{Fig 3: System two RFC}
\end{figure}

To investigate and characterize the behaviour of RFC model dynamics under a fault in the railway grid, the following cases have been studied
\begin{itemize}
	\item Case 1: A single RFC feeding a catenary section (Single feeding mode), see \Cref{Fig 2: System one RFC}, is simulated to understand the dynamic behaviour of an RFC by itself.
	\item Case 2: Two RFCs feeding a catenary section (Interconnected mode), see \Cref{Fig 3: System two RFC}, are simulated to understand the dynamic interactions between a pair RFCs, one in each converter station.
\end{itemize}

%The two cases are chosen due to their simplicity and to obtain basic understanding of the RFC dynamics.

%For simplicity, the step-down transformer reactance is included in the motor reactance and the step-up transformer reactance is included in the generator reactance.

The systems are investigated during no-load situation to avoid influence from the load. The system is disturbed by applying a balanced fault at 10 km from RFC 1. The fault is initiated at 1.8~seconds and cleared after 200~ms.

One of the most common RFC types used in Sweden is Q48/Q49. This type of RFC has been chosen for the studies  to investigate the RFC model behaviour. Data about Q48/Q49 is given in \Cref{Tab 1: Motor Parameters}. An equivalent impedance of the Booster Transformer (BT) catenary based on \cite{Fridman2006} is used, see \Cref{Tab 3: System Parameters}.

The pre-fault conditions are obtained by a load flow calculation using parts of the software Train Power System Simulator (TPSS) presented in \cite{Abrahamsson117}. TPSS uses GAMS \cite{gamsweb} and MatLab to model and solve load flows in phasor domain for low-frequency railway power systems synchronously connected to the public grid.

%\begin{table}[h!]
%	\begin{center}
%		\begin{tabular}{l |c } 		
%			\hline
%			$X^m_{q},\,X^m_{d},\,X'^m_{d},\, X''^m_{q},\, X''^m_{d}$ &  0.49, 1.02, 0.3, 0.3, 0.21 [p.u.]\\
%			$T'^{m}_{do},\,T''^{m}_{do},\,T''^{m}_{qo}$ & 3.6, 0.04, 0.09 [s]\\
%			Inertia constant motor, $H_M$ & 1.06 [MWs/MVA]\\
%			Rated power motor & 10.7~[MVA]\\
%			Transformer ratio motor & 80~[kV]/6.3~[kV] \\
%			Transformer leakage reactance motor & 7.9\%\\
%			$X^g_{q},\,X^g_{d},\,X'^g_{d},\, X''^g_{q},\, X''^g_{d}$ &  0.53, 1.39, 0.16, 0.10, 0.12 [p.u.]\\
%			$T'^{g}_{do},\,T''^{g}_{do},\,T''^{g}_{qo}$ & 11.2, 0.07, 4 [s]\\
%			Inertia constant generator, $H_G$ & 1.14 [MWs/MVA]\\
%			Rated power generator & 10~[MVA]\\
%			Transformer ratio generator& 5.2~[kV]/17~[kV] \\
%			Transformer leakage reactance generator & 4.2\%\\
%			\hline
%		\end{tabular}
%		\caption{RFC parameters of the Q48/Q49.} \label{Tab 1: Motor Parameters}
%	\end{center}
%\end{table}

\begin{table}[h!]
	\begin{center}
		\begin{tabular}{l |c } 		
			\hline
			$X^m_{q},\,X^m_{d},\,X'^m_{d},\, X''^m_{q},\, X''^m_{d}$ &  0.49, 1.02, 0.3, 0.3, 0.21 [p.u.]\\
			$T'^{m}_{do},\,T''^{m}_{do},\,T''^{m}_{qo}$ & 3.6, 0.04, 0.09 [s]\\
			Inertia constant motor, $H_M$ & 1.06 [MWs/MVA]\\
			Rated power motor & 10.7~[MVA]\\
			Transformer ratio motor & 80~[kV]/6.3~[kV] \\
			\begin{tabular}{@{}l@{}} Transformer leakage reactance \\ of the motor, $X_T^m$ \end{tabular} & 7.9\%\\
			$X^g_{q},\,X^g_{d},\,X'^g_{d},\, X''^g_{q},\, X''^g_{d}$ &  0.53, 1.39, 0.16, 0.10, 0.12 [p.u.]\\
			$T'^{g}_{do},\,T''^{g}_{do},\,T''^{g}_{qo}$ & 11.2, 0.07, 4 [s]\\
			Inertia constant generator, $H_G$ & 1.14 [MWs/MVA]\\
			Rated power generator & 10~[MVA]\\
			Transformer ratio generator& 5.2~[kV]/17~[kV] \\
			\begin{tabular}{@{}l@{}} Transformer leakage reactance \\ of the generator, $X_T^g$  \end{tabular}  & 4.2\%\\
			\hline
		\end{tabular}
		\caption{RFC parameters of the Q48/Q49.} \label{Tab 1: Motor Parameters}
	\end{center}
\end{table}

\begin{table}[h!]
	\begin{center}
		\begin{tabular}{l |c } 		
			\hline
			Catenary impedance &  0.2+j0.2~[$\Omega$/km]\\
			Base power, $S_B$ & 10~[MVA]\\
			Base voltage railway side, $U_B$& 16.5~[kV] \\
			Base voltage 50~Hz side, $U_{B50}$ & 6.3~[kV] \\
			\hline
		\end{tabular}
		\caption{System parameters.} \label{Tab 3: System Parameters}
	\end{center}
\end{table}

%RFCs are connected to infinite buses at the motor side, therefore the motor voltage is kept at one p.u. and reactive power from the motor is zero in steady state.
Due to the connection to the infinite bus, the excitation system on the motor has not been implemented and field voltages of the motors are kept constant. The majority of the RFC generators and motors in the Swedish railway system are equipped with brushless excitation systems. The AC5A \cite{IEEE_AVR_2005} excitation model of such an excitation system provided by MatLab Simulink has been used.

The model presented of an RFC in \Cref{Sec: Models} and its interface to the public grid and the railway grid represented in \Cref{Sec: Model2} has been implemented and simulated in MatLab Simulink, where the RFC dynamics and the electrical networks equations are solved. The fixed step solver \textit{ode4} (Runge-Kutta algorithm) is used, with a step size equal to 1~ms.

Oscillation frequencies of various power system variables are determined for both of the cases described above in this section. Oscillation frequencies are presented together with other important simulation results in \Cref{Sec: Results}. An oscillation frequency is a simple measure to quantify a dynamic behaviour of a machine, an other component, or a system as a whole. By studying oscillation frequencies, one may get a rough picture of RFC and power system transient behaviour.

To estimate the average frequency of an oscillation, an average of four to ten consecutive estimated values of the oscillation frequency $f_{osc}$ is used. The oscillation frequency $f_{osc}$ between two peaks of a graph obtained from the simulations,
\begin{equation}
f _{osc} \approx \frac{1}{\Delta t} \label{Eq 22: Eigenvalues},
\end{equation}
where $\Delta t $ is the time difference between the peaks of the graph. The peaks are found by using the \textit{Peak Finder} tool in MatLab Simulink.

%\newpage

%\input{Sections/SimulationsOLD}

\section{Results} \label{Sec: Results}

%To estimate the most dominant frequency and damping of an oscillation in a non-linear system, \Cref{Eq 22: Eigenvalues} can be used according to \cite{Danielsen2010}. \Cref{Eq 22: Eigenvalues} is based on the time domain expression of a damped sinusoidal oscillation and provides a simple characterization of such an oscillation. However, \Cref{Eq 22: Eigenvalues} cannot be used for more complex oscillations and other methods have to be used in those cases \cite{Danielsen2010,bollenyu2006}.
%To estimate the most dominant frequency oscillation that occurs when , \Cref{Eq 22: Eigenvalues}
%
%For estimating the most dominant oscillation frequency that the RFC model generates in powers, rotor speed and voltage magnitude
%
%The most dominant oscillation frequency $f_{osc}$ of active power, reactive power, voltage magntidue and rot the imaginary part of \Cref{Eq 22: Eigenvalues} is used.
%
%\begin{equation}
%\alpha \pm j\omega_{osc} \approx \frac{1}{(t_2-t_1)}\ln(\frac{y_2-y_{\infty}}{y_1-y_{\infty}})  \pm j\frac{1}{(t_2 - t_1)} \label{Eq 22: Eigenvalues}.
%\end{equation}
%
%%The approximated oscillation frequency $j\mu$ is obtained by calculating the time difference $t_2 - t_1$ between two peaks of an oscillation.
%
%The peaks are found by using the \textit{Peak Finder} tool supplied by MatLab Simulink. Depending on the signal, an average of four to ten estimated values of the oscillations frequency $f_{osc}$, over four to ten consecutive time differences is used to estimate the dominant oscillation frequency.

\subsection{Case 1}
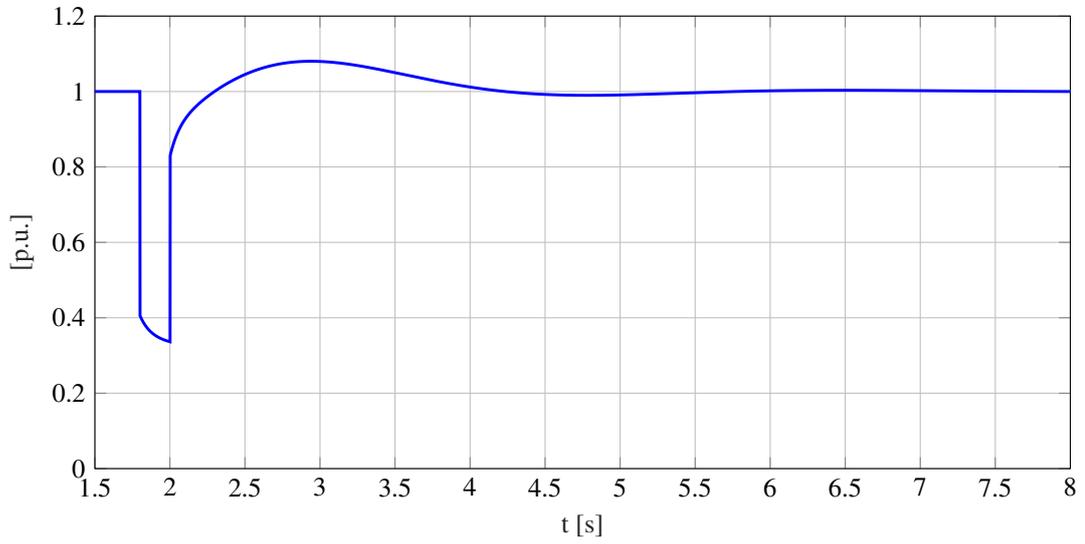
\begin{figure}
	\centering
	\setlength\fheight{6cm}
	\setlength\fwidth{\textwidth}
\begin{tikzpicture}

\begin{axis}[%
width=0.951\fwidth,
height=\fheight,
at={(0\fwidth,0\fheight)},
scale only axis,
xmin=1.5,
xmax=8,
xlabel style={font=\color{white!15!black}},
xlabel={t [s]},
ymin=0,
ymax=1.2,
ylabel style={font=\color{white!15!black}},
ylabel={[p.u.]},
axis background/.style={fill=white},
xmajorgrids,
ymajorgrids,
legend style={legend cell align=left, align=left, draw=white!15!black}
]
\addplot [color=blue, line width=1.1pt]
  table[row sep=crcr]{%
1.499	1\\
1.8	1\\
1.801	0.405373350892479\\
1.807	0.400087239390716\\
1.813	0.395214656357801\\
1.82	0.390002513201901\\
1.827	0.385248638491543\\
1.834	0.380905891926648\\
1.841	0.376932649042082\\
1.849	0.372797264073808\\
1.857	0.36904813354422\\
1.866	0.365239126853034\\
1.875	0.36181194969198\\
1.885	0.358395891377571\\
1.895	0.355339848128907\\
1.906	0.352338522417652\\
1.918	0.3494351164954\\
1.931	0.346662058728114\\
1.945	0.344041785353888\\
1.96	0.341588067221723\\
1.977	0.339175309729583\\
1.995	0.336973149389154\\
2.001	0.336308418717497\\
2.002	0.830615814896793\\
2.008	0.83991545725288\\
2.014	0.848555259479078\\
2.02	0.856591701928759\\
2.026	0.864076579523648\\
2.032	0.871057387238206\\
2.038	0.877577673973555\\
2.045	0.884655708196997\\
2.052	0.891218858823301\\
2.059	0.89731928112557\\
2.066	0.90300409079653\\
2.074	0.909046503092432\\
2.082	0.914658499469322\\
2.091	0.92052042313165\\
2.1	0.925965310753297\\
2.11	0.93159289155137\\
2.12	0.936837969795704\\
2.131	0.942230315467565\\
2.143	0.947730419315986\\
2.156	0.95331045071476\\
2.17	0.958952574646569\\
2.186	0.965014903101373\\
2.203	0.97108410833132\\
2.222	0.977493183294754\\
2.242	0.983883515585706\\
2.264	0.990554994104658\\
2.288	0.997461180405926\\
2.313	1.00428605988837\\
2.339	1.01101572271249\\
2.365	1.01739005214394\\
2.391	1.02342148118175\\
2.418	1.0293302593674\\
2.445	1.03488345625594\\
2.472	1.04008549438464\\
2.499	1.04494031404756\\
2.526	1.04945191251509\\
2.553	1.05362464466771\\
2.581	1.05759920917619\\
2.609	1.06122067260664\\
2.637	1.06449581998666\\
2.665	1.06743211349313\\
2.694	1.07012470121833\\
2.723	1.07247216697536\\
2.753	1.07454836212074\\
2.783	1.07627825969936\\
2.814	1.07771538015698\\
2.845	1.07881058300442\\
2.877	1.07959825760789\\
2.91	1.08006334856668\\
2.944	1.08019347258835\\
2.979	1.07997917317325\\
3.015	1.07941415217474\\
3.052	1.07849547261198\\
3.091	1.07718603329136\\
3.132	1.07546616807063\\
3.176	1.07326984567131\\
3.223	1.07057153581117\\
3.274	1.0672915490954\\
3.33	1.06334258891783\\
3.395	1.05840704862738\\
3.477	1.05181685815568\\
3.786	1.0266327261548\\
3.861	1.02106397653757\\
3.93	1.01627954002247\\
3.996	1.01204706962812\\
4.059	1.00834583760045\\
4.121	1.00504199483903\\
4.183	1.00208253952807\\
4.245	0.999470022142292\\
4.307	0.99720176631293\\
4.37	0.995242040352476\\
4.434	0.993595229864026\\
4.5	0.99224296931189\\
4.568	0.991194859658878\\
4.639	0.990445944465561\\
4.714	0.990002332660413\\
4.794	0.989877511834923\\
4.881	0.990091819283558\\
4.978	0.990683013510344\\
5.091	0.991727526521244\\
5.236	0.993432173935222\\
5.76	0.999898355344804\\
5.913	1.00126275581951\\
6.061	1.00224005874665\\
6.212	1.00289118650135\\
6.372	1.00323225205093\\
6.549	1.00325958559863\\
6.762	1.00293485839064\\
7.078	1.00206977286295\\
7.622	1.00058899895682\\
7.954	1.00006532898466\\
8.001	1.00001818446181\\
};
%\addlegendentry{$\text{U}_{\text{RFC 1}}$}

\end{axis}
\end{tikzpicture}%
	\caption{Voltage after the step-up transformer, railway side, Case 1.} \label{Fig 4: Case 1 Voltage}
\end{figure}

\begin{figure}
	\centering
	\setlength\fheight{7cm}
	\setlength\fwidth{\textwidth}
	\input{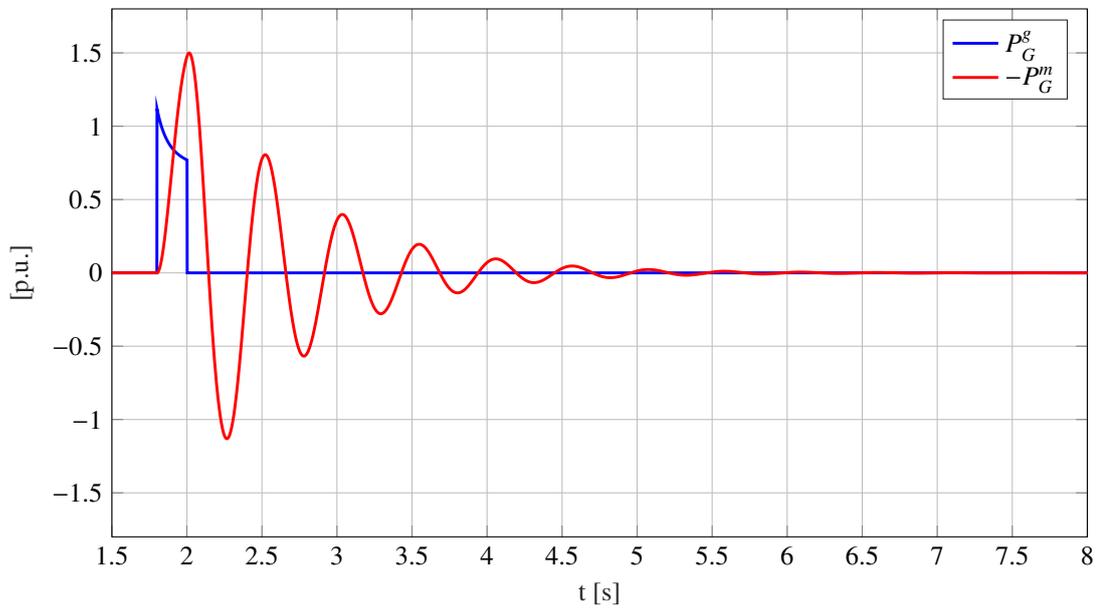}
	\caption{Generated ($P^g_G$) vs consumed ($-P^m_G$) active power of the RFC generator and motor, respectively,  Case 1.} \label{Fig 5: Case 1 AC Power}
\end{figure}

\begin{figure}
	\centering
	\setlength\fheight{6cm}
	\setlength\fwidth{\textwidth}
	\input{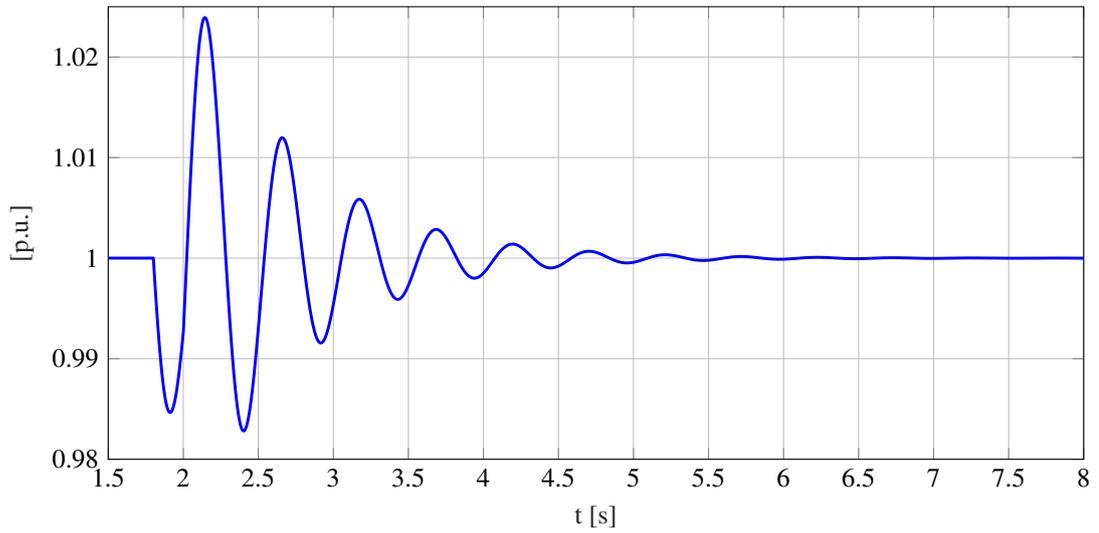}
	\caption{Rotor speed of the RFC, Case 1.} \label{Fig 6: Case 1 rotor speeds}
\end{figure}

%In single feeding mode, the RFC supplies what is required from the railway grid.
%Furthermore, no induced voltage will occur in the damper windings as induced voltage and railway grid voltage vector have equal speeds.
At fault initiation, the voltage at the railway side drops to approximately 0.40 p.u. and decays further during the fault, see \Cref{Fig 4: Case 1 Voltage}. The impedance seen from the RFC to the fault location has an X/R ratio equal to one. Therefore the RFC generator will supply both active and reactive power during the fault. As active power increases suddenly at fault initiation, the RFC motor will see an increase in load. However, due to the inertia of the RFC machine the rotor speed and the three-phase active power will increase gradually as seen in \Cref{Fig 5: Case 1 AC Power,Fig 6: Case 1 rotor speeds}.

% as energy has to be dissipated.

%Based on the simulations, the dominant oscillation frequency of the rotor is 1.96~Hz. The three-phase power of the motor and the relative rotor angle oscillates with the same frequency against the 50~Hz grid.

Based on the simulations, the average rotor oscillations is 1.96~Hz. The three-phase power of the motor and rotor angle oscillates with the same frequency against the 50~Hz grid. This type of oscillations can be expected for single synchronous machine at a power station which swing against a large power system according to \cite{Basler2005a,Rogers2000}.
On the single phase side, as there is no other machines or infinite bus, the induced voltage vector of the RFC generator and the railway grid voltage vector have the same speed.

%The low-frequency oscillation in single-feeding mode produced by the RFC model is of interest as it has been shown by \cite{Danielsen2010} that the low-frequency rotor oscillations may cause the control system of the modern locomotives or EMU to react. The rotor oscillations may get amplified due the low-frequency dynamics of the locomotive resulting in system instability.

\begin{figure}
	\centering
	\setlength\fheight{7cm}
	\setlength\fwidth{\textwidth}
	\input{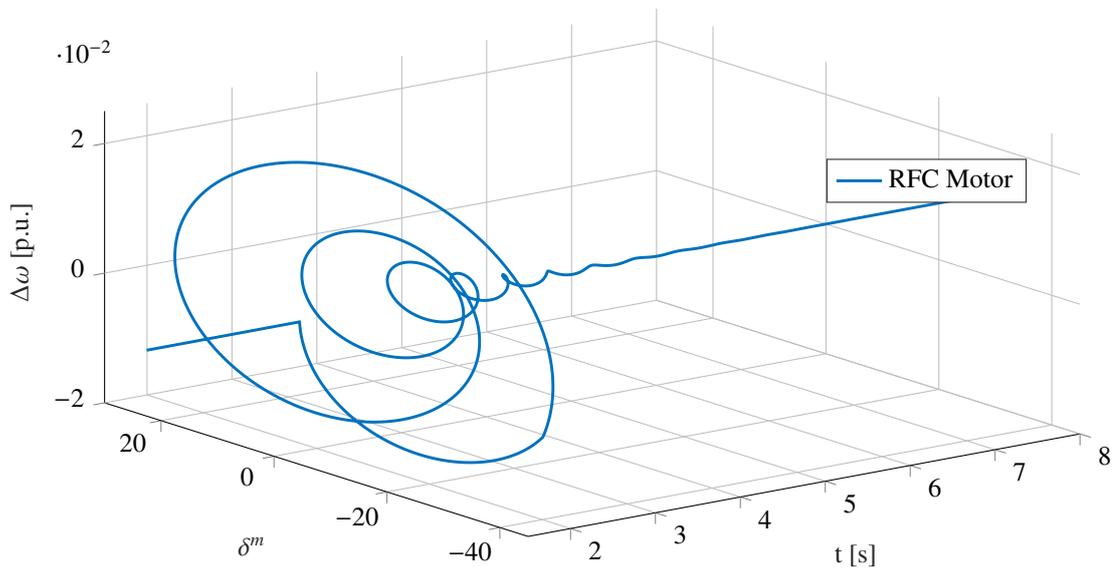}
	\caption{Phase plot $\delta^m$ vs $\Delta\omega_{p.u.} $ vs time,  Case 1.} \label{Fig 7: Case 1 Phase plot}
\end{figure}

In \Cref{Fig 7: Case 1 Phase plot}, the relative rotor angle is plotted against time and rotor speed oscillation. Both relative angle and rotor speed oscillation goes to zero, showing that the system is stable.

\subsection{Case 2}
%The RFC motors are connected each to their own infinite bus; they are only interconnected through the catenary system, i.e interconnected mode, see \Cref{Fig 3: System two RFC}.

\begin{figure*}[h!]
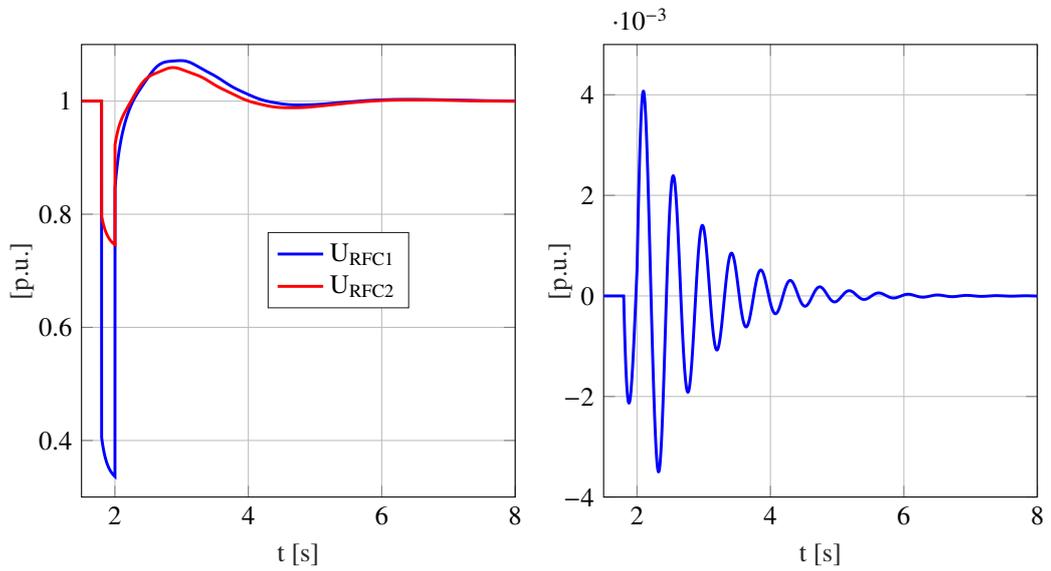
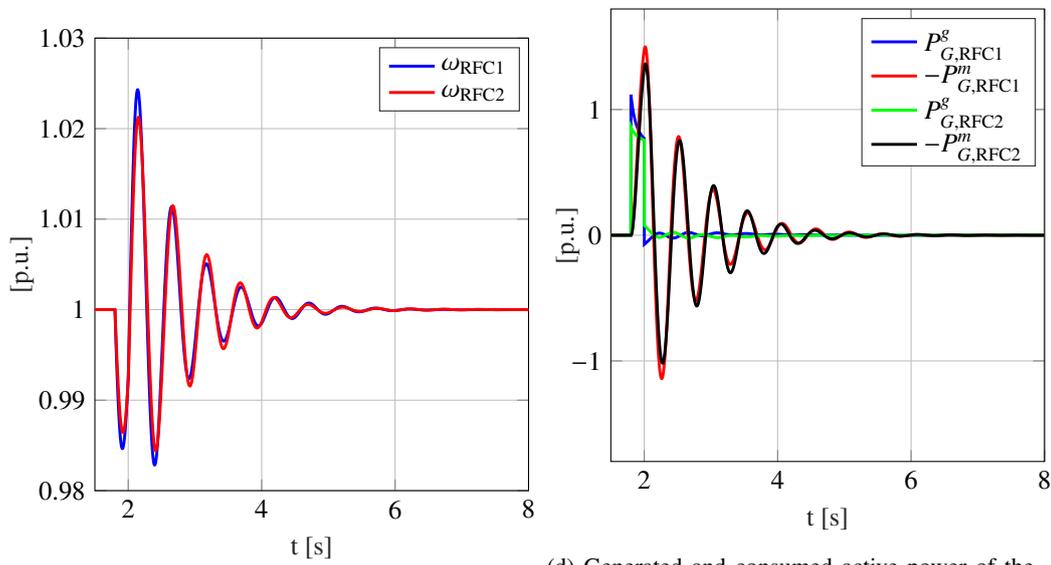

	%\centering
	\begin{subfigure}[b]{0.475\textwidth}
		\centering
		\setlength\fheight{6cm}
		\setlength\fwidth{6cm}
\begin{tikzpicture}

\begin{axis}[%
width=0.951\fwidth,
height=\fheight,
at={(0\fwidth,0\fheight)},
scale only axis,
xmin=1.5,
xmax=8,
xlabel style={font=\color{white!15!black}},
xlabel={t [s]},
ymin=0.3,
ymax=1.1,
ylabel style={font=\color{white!15!black}},
ylabel={[p.u.]},
ylabel style={yshift=-5pt},
axis background/.style={fill=white},
xmajorgrids,
ymajorgrids,
legend style={at={(0.429,0.417)}, anchor=south west, legend cell align=left, align=left, draw=white!15!black}
]
\addplot [color=blue, line width=1.1pt]
  table[row sep=crcr]{%
1.499	1\\
1.8	1\\
1.801	0.405373350892479\\
1.806	0.400938462235828\\
1.811	0.396795314451264\\
1.816	0.392921494118532\\
1.821	0.389296494568113\\
1.827	0.385248638491543\\
1.833	0.38150274630364\\
1.839	0.378032385337642\\
1.845	0.374813791732425\\
1.851	0.371825529638631\\
1.858	0.36860446772385\\
1.865	0.365642586096\\
1.872	0.362914898193912\\
1.88	0.360055708456487\\
1.888	0.357443541801276\\
1.897	0.354767609144353\\
1.906	0.352338522417652\\
1.916	0.349894589945205\\
1.926	0.347686691748265\\
1.937	0.345496264118262\\
1.949	0.343354280026919\\
1.962	0.341285255186939\\
1.976	0.33930765070882\\
1.991	0.337434608507564\\
2.001	0.336308418717497\\
2.002	0.845793688873984\\
2.007	0.853014583969522\\
2.012	0.85981739744299\\
2.017	0.866233586323055\\
2.022	0.872292249750791\\
2.027	0.878020294031971\\
2.032	0.883442586632189\\
2.038	0.889577964624472\\
2.044	0.895342417201201\\
2.05	0.900768967521209\\
2.056	0.905887574873589\\
2.062	0.910725395901229\\
2.069	0.916047318516323\\
2.076	0.921055505310132\\
2.083	0.925781132309776\\
2.09	0.930251893712546\\
2.098	0.935080556287048\\
2.106	0.939640010933086\\
2.114	0.943957437166178\\
2.123	0.948554065276237\\
2.132	0.952901835058531\\
2.142	0.95746951025729\\
2.152	0.961786396505726\\
2.163	0.966272432801059\\
2.174	0.970507979147266\\
2.186	0.974867864743588\\
2.198	0.978978336939672\\
2.21	0.98285926471598\\
2.223	0.986824474111272\\
2.236	0.990559731793997\\
2.25	0.994345090246179\\
2.265	0.998150307996985\\
2.28	1.00172000670634\\
2.296	1.00529558984906\\
2.313	1.00886590759016\\
2.332	1.01262437868123\\
2.354	1.01674234457264\\
2.383	1.02192574479276\\
2.45	1.03358118543152\\
2.511	1.04407060706838\\
2.54	1.04881741199221\\
2.563	1.05235856187598\\
2.584	1.0553620397279\\
2.603	1.05785551416885\\
2.621	1.05999950856035\\
2.639	1.06191760353051\\
2.657	1.06360351129861\\
2.675	1.06505799254858\\
2.693	1.06628852671655\\
2.712	1.0673594841403\\
2.732	1.06825786752578\\
2.754	1.06901212700317\\
2.779	1.06963037138005\\
2.809	1.07013493363239\\
2.855	1.07064939960888\\
2.951	1.07146219179092\\
2.988	1.07155556083879\\
3.017	1.07141081286434\\
3.043	1.07106295867976\\
3.068	1.07050451882516\\
3.093	1.06971396575786\\
3.118	1.06869435072603\\
3.145	1.06735668683897\\
3.175	1.06562919789644\\
3.212	1.06325118740156\\
3.375	1.05247189349254\\
3.483	1.04569670211866\\
3.528	1.04258946220856\\
3.575	1.03910332130953\\
3.654	1.032955078849\\
3.71	1.02872053044482\\
3.753	1.0256998522999\\
3.795	1.02298314366142\\
3.841	1.0202453065502\\
3.901	1.01691985288515\\
4.028	1.01016646804177\\
4.095	1.00678662959313\\
4.143	1.00459009574807\\
4.186	1.00284758811207\\
4.229	1.00133646542992\\
4.275	0.999956808534524\\
4.328	0.998604702554374\\
4.398	0.997063978190679\\
4.481	0.995461811813179\\
4.547	0.994403132396011\\
4.602	0.993743983683684\\
4.655	0.993337450634424\\
4.711	0.993139343078658\\
4.78	0.993136496098671\\
4.889	0.993390268779375\\
5.003	0.993853505714773\\
5.086	0.994418032520274\\
5.179	0.995294603818154\\
5.455	0.998035243824893\\
5.77	1.00084905940463\\
5.892	1.00159317427041\\
6.065	1.00240534846269\\
6.193	1.00279488068444\\
6.324	1.00295236584506\\
6.518	1.00293597633607\\
6.689	1.00270826715438\\
7.003	1.00197912615778\\
7.604	1.00058972824944\\
7.88	1.00021698129931\\
8.001	1.00011426699632\\
};
\addlegendentry{$\text{U}_{\text{RFC1}}$}

\addplot [color=red, line width=1.1pt]
  table[row sep=crcr]{%
1.499	1\\
1.8	1\\
1.801	0.796015772042809\\
1.807	0.792483720348454\\
1.813	0.789197940303044\\
1.82	0.785647894643164\\
1.827	0.782374677429164\\
1.834	0.779351752980199\\
1.841	0.776555555536001\\
1.849	0.773610787101553\\
1.857	0.770907362222056\\
1.866	0.768124668744582\\
1.875	0.765587391678364\\
1.885	0.763025122375506\\
1.895	0.760704557675012\\
1.906	0.758401048151585\\
1.917	0.756331180906137\\
1.929	0.754311936873755\\
1.941	0.752517172736317\\
1.954	0.750801802222608\\
1.968	0.7491962546826\\
1.982	0.747819524298871\\
1.997	0.746577536522457\\
2.001	0.746284691572447\\
2.002	0.92251982945748\\
2.008	0.926876195998458\\
2.014	0.930936536430108\\
2.02	0.934725141505728\\
2.026	0.938264353969267\\
2.033	0.942105574599685\\
2.04	0.945665160679027\\
2.047	0.948970318620189\\
2.054	0.952045854372226\\
2.062	0.955308563850879\\
2.07	0.958331683428469\\
2.079	0.961480960618948\\
2.088	0.96439818439878\\
2.098	0.967406595779897\\
2.109	0.970478115360846\\
2.121	0.973597438261343\\
2.135	0.976998676807648\\
2.152	0.980879006260748\\
2.174	0.985644369255787\\
2.211	0.993380744021419\\
2.331	1.01817825483572\\
2.354	1.02260489031933\\
2.374	1.02622772065909\\
2.392	1.02926841278653\\
2.41	1.03207498550231\\
2.427	1.03449689138696\\
2.444	1.03669173727984\\
2.461	1.03866234491414\\
2.479	1.0405150603433\\
2.498	1.04222961913943\\
2.518	1.04379758162914\\
2.54	1.04528822340456\\
2.566	1.04681123340463\\
2.602	1.04866630525791\\
2.779	1.05728141469733\\
2.805	1.0581925583815\\
2.829	1.05880255262316\\
2.851	1.05913659699229\\
2.873	1.05924014571202\\
2.895	1.05911132470237\\
2.918	1.0587400758061\\
2.942	1.05812254835215\\
2.969	1.05719842446087\\
3.003	1.05579496194799\\
3.178	1.04821229156701\\
3.233	1.04586137314008\\
3.27	1.04405351445952\\
3.304	1.04216412378992\\
3.338	1.04004460648064\\
3.376	1.03743913741843\\
3.43	1.03347986009389\\
3.501	1.02831247118625\\
3.544	1.02542220013818\\
3.588	1.02269876727032\\
3.643	1.01954258691017\\
3.795	1.01093794621715\\
3.89	1.00550466834995\\
3.936	1.00312032774545\\
3.977	1.00122039083195\\
4.018	0.999546132987357\\
4.063	0.997938998857688\\
4.119	0.99618153580092\\
4.201	0.993860921152651\\
4.284	0.991715700957181\\
4.342	0.990437766485208\\
4.391	0.989582605526063\\
4.439	0.988975142158226\\
4.49	0.988565174904739\\
4.551	0.988317449726157\\
4.638	0.988214317166367\\
4.743	0.988304904089746\\
4.817	0.988589201249118\\
4.885	0.989083428843735\\
4.963	0.989892342641424\\
5.238	0.993175652798246\\
5.387	0.995104498862414\\
5.535	0.996974506566303\\
5.651	0.998184627371444\\
5.826	0.99975610682216\\
5.947	1.00064266811592\\
6.057	1.00121301571805\\
6.199	1.00170200211974\\
6.354	1.00202252081725\\
6.49	1.00207151017544\\
6.691	1.00188114836511\\
6.92	1.00146230295769\\
7.559	1.00010864077774\\
7.871	0.999740693202233\\
8.001	0.999656859929505\\
};
\addlegendentry{$\text{U}_{\text{RFC2}}$}

\end{axis}
\end{tikzpicture}%
		\caption{Voltage after the step-up transformer, railway side.} \label{Fig 8: Case 2 Voltage}
	\end{subfigure}
	\hfill
	\begin{subfigure}[b]{0.475\textwidth}
		\centering
		\setlength\fheight{6cm}
		\setlength\fwidth{6cm}
		\input{Case2_RotorSpeed_relative_Gen.tikz}
		\caption{Relative rotor speed between the RFCs. \phantom{adaddad}} \label{Fig 10: Case 2 relative rotor speeds}
	\end{subfigure}
	\vskip\baselineskip
	\begin{subfigure}[b]{0.475\textwidth}
		\centering
		\setlength\fheight{6cm}
		\setlength\fwidth{6cm}
		\input{Case2_RotorSpeed.tikz}
		\caption{Rotor speeds of the RFCs. \phantom{asadsdasdadasdad}} \label{Fig 9: Case 2 Rotor speeds}
	\end{subfigure}
	\qquad
	\begin{subfigure}[b]{0.475\textwidth}
		\centering
		\setlength\fheight{6cm}
		\setlength\fwidth{6cm}
		\input{Case2_1ACvs3AC.tikz}
		\caption{Generated and consumed active power of the RFCs generator \& motor.} \label{Fig 11: Active3vs1}
	\end{subfigure}
	\caption[]
	{\small Case 2 results.} \label{Fig 12: Case 2 part 1}
\end{figure*}

During the fault, the rotor speed of RFC 1 drops more than the speed of RFC 2, as the fault current provided by RFC 1 generator is higher than RFC 2, see \Cref{Fig 9: Case 2 Rotor speeds}.

The simulations show that the rotor speed oscillation of RFC 1 and RFC 2 has an average frequency of 1.96~Hz as seen in \Cref{Fig 9: Case 2 Rotor speeds}. The three-phase powers will oscillate at 1.96~Hz for both RFC, see \Cref{Fig 11: Active3vs1}. This is in line with Case 1, as the both RFCs are of the same type.

\begin{figure*}[h!]
	%\centering
	\begin{subfigure}[b]{0.5\textwidth}
		\centering
		\setlength\fheight{6cm}
		\setlength\fwidth{6cm}
		\input{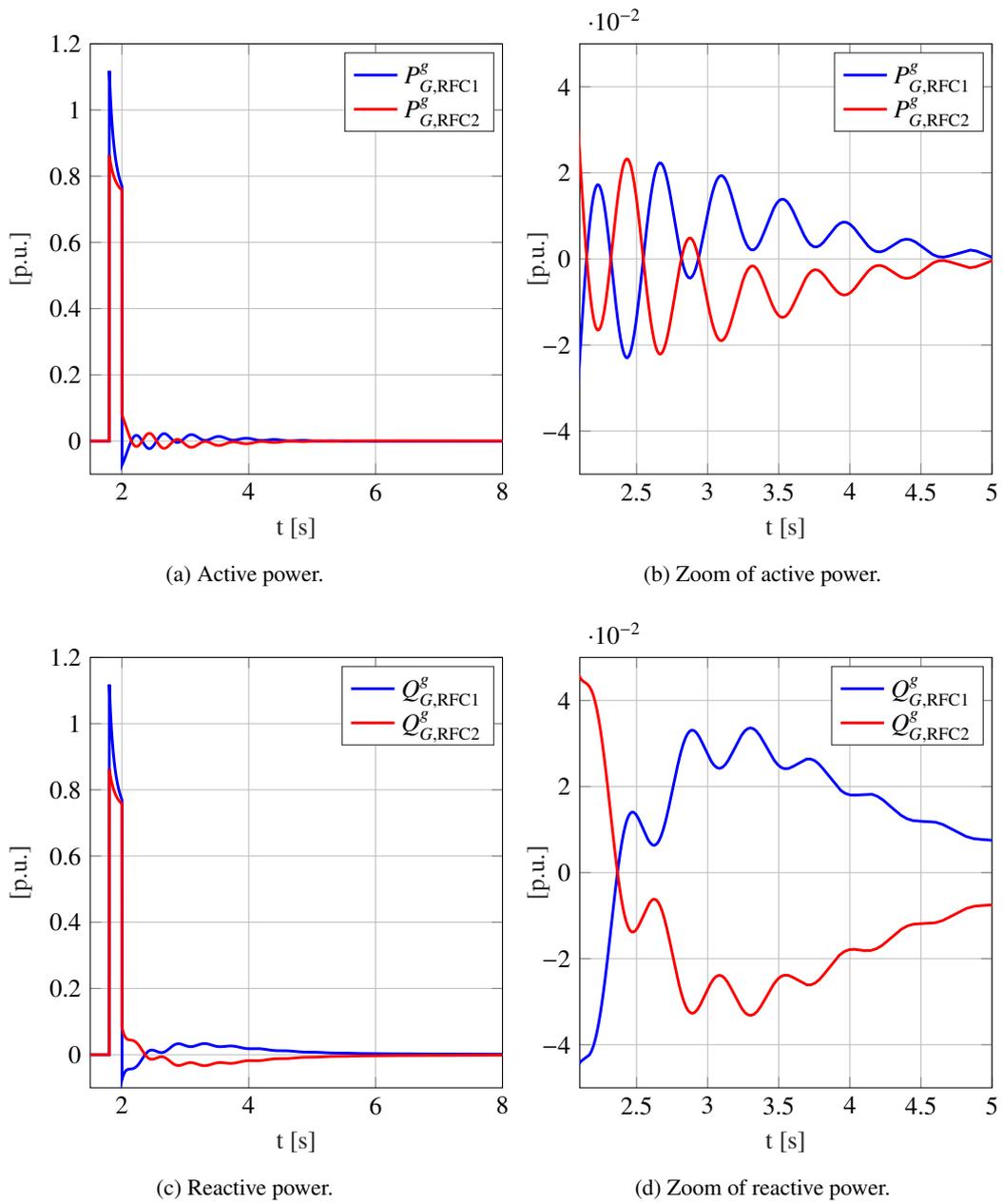}
		\caption{Active power.}  \label{Fig 13: 1 phase Ac Power}
	\end{subfigure}
	\quad
	\begin{subfigure}[b]{0.5\textwidth}
		\centering
		\setlength\fheight{6cm}
		\setlength\fwidth{6cm}
\begin{tikzpicture}

\begin{axis}[%
width=0.951\fwidth,
height=\fheight,
at={(0\fwidth,0\fheight)},
scale only axis,
xmin=2.1,
xmax=5,
xlabel style={font=\color{white!15!black}},
xlabel={t [s]},
ymin=-0.05,
ymax=0.05,
ylabel style={font=\color{white!15!black}},
ylabel={[p.u.]},
ylabel style={yshift=-10pt},
axis background/.style={fill=white},
xmajorgrids,
ymajorgrids,
legend style={legend cell align=left, align=left, draw=white!15!black}
]
\addplot [color=blue, line width=1.1pt]
  table[row sep=crcr]{%
2.085	-0.0311839698133056\\
2.111	-0.0173108315987509\\
2.125	-0.0102453077775948\\
2.137	-0.00459621037949631\\
2.148	0.00015365740483908\\
2.158	0.00405132714201528\\
2.167	0.00717600197184254\\
2.176	0.00990800946652648\\
2.185	0.012225158240005\\
2.193	0.0139235656127816\\
2.201	0.0152744680248329\\
2.209	0.0162745833228328\\
2.217	0.0169242780362797\\
2.225	0.0172274331458215\\
2.233	0.0171913139031492\\
2.242	0.0167583755391245\\
2.251	0.0159292185978916\\
2.26	0.01472848350177\\
2.27	0.0129943776689139\\
2.28	0.0108846368510687\\
2.291	0.00819376308614928\\
2.304	0.0046193756598818\\
2.321	-0.000469512124531946\\
2.355	-0.0107418629416145\\
2.368	-0.0142398780129618\\
2.379	-0.0168483857044031\\
2.389	-0.0188757950314162\\
2.399	-0.020529485398157\\
2.408	-0.0216694267995212\\
2.417	-0.0224603469941993\\
2.426	-0.0228910304908361\\
2.435	-0.0229570645469721\\
2.444	-0.0226605848458021\\
2.453	-0.0220098959819666\\
2.463	-0.0208886927167615\\
2.473	-0.0193742875209928\\
2.483	-0.0175003321680069\\
2.494	-0.015071234095446\\
2.506	-0.0120506119381592\\
2.52	-0.00814979502685098\\
2.572	0.00744711683383414\\
2.587	0.0115233329491087\\
2.6	0.0146591130803628\\
2.611	0.0169550065750723\\
2.622	0.0188768988215173\\
2.632	0.0202716228190098\\
2.642	0.0213131401715572\\
2.652	0.0219924992959761\\
2.662	0.0223085966501682\\
2.672	0.0222679779800323\\
2.683	0.0218279648286224\\
2.694	0.0210008442781682\\
2.706	0.019699961195851\\
2.719	0.017886283860209\\
2.733	0.0155566826533224\\
2.751	0.0121698815817162\\
2.802	0.00233174511637557\\
2.817	-5.14543508263188e-05\\
2.83	-0.00176062293901502\\
2.842	-0.00299337571033576\\
2.854	-0.00386167535502402\\
2.866	-0.00434530780357179\\
2.878	-0.00443648758382587\\
2.89	-0.00413989812878413\\
2.902	-0.00347228779082265\\
2.915	-0.00236305449561769\\
2.929	-0.000774612428640964\\
2.945	0.0014452809686567\\
2.965	0.00463353077506845\\
3.018	0.0133081620879008\\
3.034	0.0154563238994836\\
3.049	0.0170990916556608\\
3.063	0.0182536940721914\\
3.076	0.0189695268044119\\
3.089	0.0193294814887786\\
3.102	0.0193340826693476\\
3.116	0.018956598576489\\
3.13	0.0182128253417648\\
3.146	0.0169716364088828\\
3.164	0.0151777466282219\\
3.188	0.0123661461549354\\
3.237	0.00652781695754001\\
3.256	0.00469579926074459\\
3.273	0.00342157432132773\\
3.289	0.00259103273734596\\
3.305	0.00214522687473284\\
3.321	0.00208951789717737\\
3.337	0.0024087888265516\\
3.354	0.00312037479658756\\
3.373	0.00429450161970379\\
3.397	0.00618031965320753\\
3.465	0.0117735743237528\\
3.485	0.0129107936956885\\
3.503	0.0135817990713321\\
3.521	0.0138822860306043\\
3.539	0.013803738847816\\
3.557	0.0133628768234484\\
3.577	0.0124968125751685\\
3.6	0.0111153242559485\\
3.633	0.00871929841281194\\
3.678	0.00550074746339035\\
3.702	0.0041626989816308\\
3.724	0.00330601719881862\\
3.745	0.00286421563688855\\
3.766	0.0027962984794403\\
3.788	0.00309696876359133\\
3.813	0.0038188652560045\\
3.847	0.00521034814898336\\
3.904	0.00756945288406286\\
3.931	0.0082815942848935\\
3.955	0.00855743428171429\\
3.979	0.00846540976592891\\
4.004	0.00799479137733172\\
4.032	0.00709115971614871\\
4.072	0.00538575502318928\\
4.128	0.00303478358614484\\
4.158	0.00215990338299576\\
4.186	0.00171277658281888\\
4.214	0.00164099208150148\\
4.245	0.00194940524016296\\
4.285	0.00275212740652009\\
4.36	0.00430988387663866\\
4.394	0.0045843553174052\\
4.426	0.00447085867179098\\
4.46	0.0039763988488577\\
4.505	0.0029144978026423\\
4.584	0.00100756313015005\\
4.622	0.000514155405573646\\
4.659	0.000415963178770795\\
4.7	0.000690888712014726\\
4.848	0.00206190527894634\\
4.891	0.00182852998085714\\
4.945	0.0011330035862116\\
5.041	-0.000145952435429919\\
};
\addlegendentry{$P^g_{G\text{,RFC1}}$}

\addplot [color=red, line width=1.1pt]
  table[row sep=crcr]{%
2.066	0.0435124397220115\\
2.103	0.0227154043815858\\
2.118	0.0147531004205614\\
2.13	0.00878374214138589\\
2.141	0.00371904756652874\\
2.151	-0.000485472999221415\\
2.16	-0.00390428116214281\\
2.169	-0.00694765276734088\\
2.178	-0.00959288554320104\\
2.187	-0.0118228770838495\\
2.195	-0.0134470131249458\\
2.203	-0.0147292545243154\\
2.211	-0.0156682274415605\\
2.219	-0.0162656348238723\\
2.227	-0.0165261402390815\\
2.236	-0.0164259920954581\\
2.245	-0.0159240182124893\\
2.254	-0.0150395886992616\\
2.263	-0.0137964494215019\\
2.273	-0.0120287934564498\\
2.283	-0.00989785143086763\\
2.294	-0.0071954879782794\\
2.307	-0.00362030099052468\\
2.325	0.00175919258427193\\
2.356	0.0110785927607893\\
2.369	0.0145810949328204\\
2.38	0.0171951613432579\\
2.39	0.0192254898257627\\
2.4	0.0208771292607626\\
2.409	0.0220089646928674\\
2.418	0.0227844652981783\\
2.427	0.0231916152194138\\
2.436	0.0232257759431231\\
2.445	0.0228894427935771\\
2.454	0.0221918357414985\\
2.463	0.0211483681071218\\
2.473	0.0196090263374238\\
2.483	0.0177046082895096\\
2.494	0.0152386519655821\\
2.506	0.0121778094477989\\
2.521	0.00794238031358852\\
2.543	0.00127953104252221\\
2.57	-0.00682808222929054\\
2.585	-0.010935740885758\\
2.598	-0.0141069098065838\\
2.61	-0.0166343118047614\\
2.621	-0.0185654061086087\\
2.632	-0.0200947009702475\\
2.642	-0.021118337209149\\
2.652	-0.0217842575228246\\
2.662	-0.022091754967736\\
2.672	-0.0220472904621012\\
2.683	-0.0216078899017891\\
2.694	-0.0207852488023832\\
2.706	-0.0194917423330878\\
2.719	-0.0176862907519411\\
2.733	-0.0153624727743953\\
2.751	-0.0119735719752381\\
2.804	-0.00170927875973881\\
2.819	0.000658722519068178\\
2.832	0.00234371755592733\\
2.844	0.00354454096844314\\
2.856	0.00437172149105436\\
2.868	0.00480572215890973\\
2.88	0.00484012422777269\\
2.892	0.0044816458703103\\
2.904	0.00374963269954165\\
2.917	0.00257125738385255\\
2.931	0.000913548095365258\\
2.947	-0.00137296803954357\\
2.968	-0.00478609645862171\\
3.014	-0.0123855097416126\\
3.031	-0.0147436400712415\\
3.046	-0.0164555291708606\\
3.06	-0.017683252360607\\
3.073	-0.018472491419649\\
3.086	-0.0189089883597724\\
3.099	-0.0189907891306689\\
3.113	-0.018694006782658\\
3.127	-0.0180248584116178\\
3.142	-0.0169412076796522\\
3.159	-0.0153351345509289\\
3.18	-0.0129589278591293\\
3.246	-0.00519176064596749\\
3.264	-0.00361478554115102\\
3.28	-0.00256916939857277\\
3.296	-0.00190040360054411\\
3.311	-0.00163199225318778\\
3.327	-0.00172535885087832\\
3.343	-0.00218776781104957\\
3.361	-0.00309628859989886\\
3.381	-0.00448869151859732\\
3.454	-0.010663364208277\\
3.476	-0.0121171704014547\\
3.495	-0.0130037581613003\\
3.513	-0.0134754710927183\\
3.531	-0.0135688187737539\\
3.549	-0.0132898036286129\\
3.568	-0.0126243820921612\\
3.589	-0.0115171899301583\\
3.616	-0.00969229633263069\\
3.688	-0.00460059015305703\\
3.711	-0.00346888236342657\\
3.732	-0.00279994036936948\\
3.753	-0.00250849528683617\\
3.774	-0.00258795925280175\\
3.797	-0.0030539955094735\\
3.824	-0.00398911219341969\\
3.874	-0.00619606146894647\\
3.909	-0.00753399401665611\\
3.935	-0.00816598093724608\\
3.959	-0.00838848865640074\\
3.983	-0.00824190299509908\\
4.008	-0.00772083160774528\\
4.037	-0.00673684647110839\\
4.082	-0.00477913521677742\\
4.129	-0.00284974992203502\\
4.159	-0.00199113954422891\\
4.187	-0.00156302836932376\\
4.215	-0.00151207300105227\\
4.246	-0.00184269753205513\\
4.287	-0.00269126169967926\\
4.358	-0.00419497380642042\\
4.392	-0.00449871000422597\\
4.424	-0.00441302708549784\\
4.458	-0.00394338432167451\\
4.501	-0.00295217874102249\\
4.587	-0.000892231165956048\\
4.624	-0.000439614500766794\\
4.661	-0.000366045616251753\\
4.703	-0.000674458922027554\\
4.842	-0.00203362986480116\\
4.885	-0.00185319081754365\\
4.936	-0.00124090395622822\\
5.047	0.000217417528187269\\
};
\addlegendentry{$P^g_{G\text{,RFC2}}$}

\end{axis}
\end{tikzpicture}%
		\caption{Zoom of active power.}  \label{Fig 14: 1 phase Ac Power zoom}
	\end{subfigure}
	\vskip\baselineskip
	\begin{subfigure}[b]{0.5\textwidth}
		\centering
		\setlength\fheight{6cm}
		\setlength\fwidth{6cm}
\begin{tikzpicture}

\begin{axis}[%
width=0.951\fwidth,
height=\fheight,
at={(0\fwidth,0\fheight)},
scale only axis,
xmin=1.5,
xmax=8,
xlabel style={font=\color{white!15!black}},
xlabel={t [s]},
ymin=-0.1,
ymax=1.2,
ylabel style={font=\color{white!15!black}},
ylabel={[p.u.]},
axis background/.style={fill=white},
xmajorgrids,
ymajorgrids,
legend style={legend cell align=left, align=left, draw=white!15!black}
]
\addplot [color=blue, line width=1.1pt]
  table[row sep=crcr]{%
1.499	0\\
1.8	0\\
1.801	1.1184544117839\\
1.806	1.09411592121583\\
1.811	1.07162038743906\\
1.816	1.05079856430269\\
1.821	1.0314991711488\\
1.826	1.01358682335786\\
1.831	0.996940207996898\\
1.836	0.981450487050402\\
1.841	0.967019897901357\\
1.846	0.953560496296173\\
1.851	0.940992815442195\\
1.857	0.926987597413268\\
1.863	0.914050717447715\\
1.869	0.902081977791516\\
1.875	0.890991805484774\\
1.881	0.88070002081812\\
1.887	0.871134761914835\\
1.893	0.862231543812008\\
1.899	0.853932433650245\\
1.905	0.846185326304008\\
1.912	0.837782377566343\\
1.919	0.830000491243231\\
1.926	0.822780579930816\\
1.933	0.816070156516819\\
1.94	0.809822535082587\\
1.948	0.803195964380368\\
1.956	0.79706477968781\\
1.964	0.791380710204454\\
1.973	0.785467871757412\\
1.982	0.780014326885532\\
1.991	0.774974020345713\\
2.001	0.769809692954921\\
2.002	-0.0750662703050402\\
2.008	-0.070872927707553\\
2.014	-0.0670699107189083\\
2.021	-0.0631067445286604\\
2.028	-0.0596279202921455\\
2.035	-0.0566035250282013\\
2.042	-0.0540015503379578\\
2.049	-0.0517885882283835\\
2.056	-0.0499303829691851\\
2.064	-0.0481966492059414\\
2.072	-0.0468292307290064\\
2.081	-0.0456649702478931\\
2.091	-0.04475233170953\\
2.103	-0.044058897990384\\
2.118	-0.043592991849728\\
2.168	-0.0423726897941741\\
2.183	-0.0414597503916045\\
2.197	-0.0402164394230997\\
2.21	-0.0386887547160644\\
2.223	-0.0367876698436707\\
2.236	-0.0345142966840637\\
2.249	-0.0318827809365967\\
2.263	-0.0286786438569955\\
2.278	-0.0248730600472449\\
2.296	-0.0199060171425671\\
2.323	-0.0119982257284565\\
2.351	-0.00388397349816927\\
2.368	0.000651757675980136\\
2.382	0.00401398543767861\\
2.395	0.00675874731069115\\
2.407	0.00892642932620724\\
2.419	0.0107166618333192\\
2.431	0.0121173391458296\\
2.443	0.013129301282202\\
2.455	0.0137655455904042\\
2.468	0.0140587522802633\\
2.481	0.0139817951624899\\
2.496	0.0135025047904307\\
2.513	0.0125675963698999\\
2.536	0.0108978064894885\\
2.578	0.00780258818925184\\
2.596	0.00687912591432926\\
2.612	0.00641687886697895\\
2.627	0.00634815304366221\\
2.642	0.00666428908177785\\
2.657	0.00737642588014964\\
2.672	0.00847623756168758\\
2.688	0.0100465187375622\\
2.705	0.0121062259646152\\
2.725	0.0149324185619157\\
2.753	0.019314517318243\\
2.791	0.0252178514232959\\
2.811	0.0279249238747354\\
2.828	0.0298547926508146\\
2.844	0.0312918915775633\\
2.859	0.0322700710826531\\
2.874	0.0328767900096008\\
2.889	0.0331161021020367\\
2.905	0.0329906355173133\\
2.922	0.0324768006995626\\
2.941	0.0315291371568556\\
2.966	0.0298876765016391\\
3.024	0.0259429319530238\\
3.045	0.024953964429077\\
3.065	0.0243875922422667\\
3.084	0.0242279703669563\\
3.103	0.0244418798612127\\
3.123	0.0250432256218218\\
3.145	0.0260795986725402\\
3.173	0.027788567135012\\
3.238	0.0318989240888836\\
3.261	0.0329039335001422\\
3.282	0.0334583100523318\\
3.302	0.033624720109481\\
3.322	0.0334325131862219\\
3.343	0.0328699664911696\\
3.367	0.0318470096831724\\
3.398	0.0301199801025671\\
3.467	0.0261387918143026\\
3.494	0.0250366256723584\\
3.519	0.0243916748384923\\
3.544	0.0241235915220024\\
3.571	0.02421650527282\\
3.604	0.0247272229950557\\
3.695	0.026366993754916\\
3.724	0.0263811939871186\\
3.752	0.0260206585031462\\
3.781	0.0252710251840114\\
3.815	0.0240040852036536\\
3.937	0.0190789728357039\\
3.969	0.0183882598326601\\
4.003	0.018037363690711\\
4.043	0.0180172526276365\\
4.156	0.0181909129620497\\
4.194	0.0177293673775356\\
4.235	0.0168447074515168\\
4.293	0.0151692709168234\\
4.36	0.0133158683220564\\
4.403	0.0125057250816969\\
4.446	0.0120708068198336\\
4.5	0.011925763215439\\
4.603	0.0117161550457183\\
4.655	0.0111656769249908\\
4.718	0.0100952566445294\\
4.818	0.00838957886218594\\
4.873	0.00787051481293233\\
4.937	0.00766594535471832\\
5.071	0.0073414478099707\\
5.145	0.00667242223184772\\
5.284	0.00533280581389306\\
5.358	0.00508701616313267\\
5.563	0.00465073723066567\\
5.771	0.00364251845036989\\
6.126	0.00296833356675741\\
6.252	0.0027999535174601\\
6.487	0.00258157844392315\\
6.695	0.00230987993520415\\
6.982	0.00207446173863879\\
7.236	0.00192474236402518\\
8.001	0.00140037095395584\\
};
\addlegendentry{$Q^g_{G\text{,RFC1}}$}

\addplot [color=red, line width=1.1pt]
  table[row sep=crcr]{%
1.499	0\\
1.8	0\\
1.801	0.862543960090314\\
1.807	0.854906446002337\\
1.813	0.847831950747073\\
1.819	0.841268928990056\\
1.826	0.834197313753723\\
1.833	0.827694453893878\\
1.84	0.821703309895762\\
1.847	0.816173696400694\\
1.855	0.810362827576823\\
1.863	0.805039552110314\\
1.871	0.800153885413971\\
1.88	0.795126811384504\\
1.889	0.790545629279322\\
1.898	0.786364839665572\\
1.908	0.782141919529149\\
1.918	0.778319856120687\\
1.929	0.774532870552557\\
1.94	0.771141485985282\\
1.952	0.767850841812313\\
1.964	0.76494931703755\\
1.977	0.762207717314523\\
1.99	0.759850763552635\\
2.001	0.758135719641613\\
2.002	0.0817466077628151\\
2.008	0.0768307791011154\\
2.014	0.072393919703682\\
2.02	0.0684082259064631\\
2.026	0.064845975668284\\
2.032	0.0616796901551382\\
2.038	0.0588822511944063\\
2.045	0.0560491905440834\\
2.052	0.0536403722716976\\
2.059	0.0516152672163859\\
2.067	0.0497200751150508\\
2.075	0.048217171966316\\
2.084	0.0469262998126005\\
2.094	0.0458995684462558\\
2.105	0.0451516642467951\\
2.119	0.0445918293263663\\
2.148	0.0439431342014664\\
2.167	0.0432907911431641\\
2.182	0.0423973616743627\\
2.195	0.0412660160663698\\
2.208	0.0397604058191412\\
2.221	0.0378612575212465\\
2.234	0.0355677192726969\\
2.247	0.0328954451561447\\
2.261	0.0296293247166837\\
2.276	0.0257451252800234\\
2.294	0.0206818626949516\\
2.323	0.0120608283263888\\
2.348	0.00476641264785016\\
2.365	0.000193488405768605\\
2.379	-0.00320202845394668\\
2.392	-0.00598598197268174\\
2.404	-0.00820100483415054\\
2.416	-0.0100518291611138\\
2.428	-0.011527258346451\\
2.44	-0.0126273169264213\\
2.452	-0.013362488617398\\
2.465	-0.013770436045629\\
2.478	-0.013809739722717\\
2.492	-0.0134922667712445\\
2.508	-0.0127569294081109\\
2.528	-0.0114445161374288\\
2.59	-0.00706460380449236\\
2.607	-0.00640101199636156\\
2.622	-0.00617395412916366\\
2.637	-0.00632867800025316\\
2.651	-0.00683704623172687\\
2.665	-0.00769755185706522\\
2.68	-0.00899216561326632\\
2.696	-0.0107555897824643\\
2.714	-0.0131294964074726\\
2.736	-0.0164229556684354\\
2.798	-0.0259430168210848\\
2.816	-0.0282187626818047\\
2.833	-0.0299897801138656\\
2.849	-0.0312684717828109\\
2.864	-0.0321002388917009\\
2.879	-0.0325706650997688\\
2.895	-0.0326859781415685\\
2.911	-0.0324332849737399\\
2.929	-0.0317700899884432\\
2.95	-0.03060641903887\\
2.981	-0.0284652640903857\\
3.02	-0.02583640517493\\
3.042	-0.0247274901775807\\
3.062	-0.0240901221549983\\
3.081	-0.0238610071147214\\
3.1	-0.0240104232790301\\
3.12	-0.0245543586796728\\
3.142	-0.0255430106713739\\
3.169	-0.0271501936519272\\
3.24	-0.0315849254453973\\
3.263	-0.0325392994127132\\
3.284	-0.0330466666648483\\
3.304	-0.0331736747965135\\
3.325	-0.0329314427960483\\
3.347	-0.032297243890854\\
3.372	-0.0311889200327915\\
3.406	-0.0292644332931928\\
3.464	-0.0259513293314715\\
3.491	-0.0248141704190719\\
3.516	-0.0241294773750731\\
3.541	-0.0238228785605994\\
3.568	-0.0238822982276243\\
3.6	-0.0243496300333987\\
3.699	-0.0260937268842252\\
3.728	-0.0260583825653224\\
3.756	-0.0256537590429566\\
3.786	-0.0248383810813966\\
3.822	-0.0234621834189603\\
3.927	-0.0191792491193823\\
3.96	-0.0183668268395696\\
3.993	-0.0179325629775118\\
4.03	-0.0178271064885855\\
4.098	-0.0181068617285369\\
4.147	-0.0180990757143586\\
4.186	-0.0177159479300997\\
4.226	-0.0169392851600403\\
4.277	-0.0155430123200642\\
4.363	-0.0131613640631869\\
4.406	-0.0123852362945556\\
4.45	-0.011975354389433\\
4.507	-0.0118575317984497\\
4.6	-0.0116747362214742\\
4.652	-0.011151109669358\\
4.714	-0.0101239826407529\\
4.82	-0.00833111997622282\\
4.875	-0.00782919105947144\\
4.94	-0.0076359918440474\\
5.069	-0.00733015911069579\\
5.142	-0.00668501937371069\\
5.287	-0.00530292068092209\\
5.362	-0.00507083337323166\\
5.556	-0.00468000716884021\\
5.782	-0.00362336346424286\\
6.071	-0.00314460176824838\\
6.214	-0.0028117193482462\\
6.963	-0.00209829417896223\\
7.215	-0.00193192491126304\\
7.818	-0.00151937184117479\\
8.001	-0.00139944909013501\\
};
\addlegendentry{$Q^g_{G\text{,RFC2}}$}

\end{axis}
\end{tikzpicture}%
		\caption{Reactive power.} \label{Fig 15: 1 phase Re Power}
	\end{subfigure}
	\quad
	\begin{subfigure}[b]{0.5\textwidth}
		\centering
		\setlength\fheight{6cm}
		\setlength\fwidth{6cm}
\begin{tikzpicture}

\begin{axis}[%
width=0.951\fwidth,
height=\fheight,
at={(0\fwidth,0\fheight)},
scale only axis,
xmin=2.1,
xmax=5,
xlabel style={font=\color{white!15!black}},
xlabel={t [s]},
ymin=-0.05,
ymax=0.05,
ylabel style={font=\color{white!15!black}},
ylabel={[p.u.]},
ylabel style={yshift=-10pt},
axis background/.style={fill=white},
xmajorgrids,
ymajorgrids,
legend style={legend cell align=left, align=left, draw=white!15!black}
]
\addplot [color=blue, line width=1.1pt]
  table[row sep=crcr]{%
1.499	0\\
1.8	0\\
1.801	1.1184544117839\\
1.806	1.09411592121583\\
1.811	1.07162038743906\\
1.816	1.05079856430269\\
1.821	1.0314991711488\\
1.826	1.01358682335786\\
1.831	0.996940207996898\\
1.836	0.981450487050402\\
1.841	0.967019897901357\\
1.846	0.953560496296173\\
1.851	0.940992815442195\\
1.857	0.926987597413268\\
1.863	0.914050717447715\\
1.869	0.902081977791516\\
1.875	0.890991805484774\\
1.881	0.88070002081812\\
1.887	0.871134761914835\\
1.893	0.862231543812008\\
1.899	0.853932433650245\\
1.905	0.846185326304008\\
1.912	0.837782377566343\\
1.919	0.830000491243231\\
1.926	0.822780579930816\\
1.933	0.816070156516819\\
1.94	0.809822535082587\\
1.948	0.803195964380368\\
1.956	0.79706477968781\\
1.964	0.791380710204454\\
1.973	0.785467871757412\\
1.982	0.780014326885532\\
1.991	0.774974020345713\\
2.001	0.769809692954921\\
2.002	-0.0750662703050402\\
2.008	-0.070872927707553\\
2.014	-0.0670699107189083\\
2.021	-0.0631067445286604\\
2.028	-0.0596279202921455\\
2.035	-0.0566035250282013\\
2.042	-0.0540015503379578\\
2.049	-0.0517885882283835\\
2.056	-0.0499303829691851\\
2.064	-0.0481966492059414\\
2.072	-0.0468292307290064\\
2.081	-0.0456649702478931\\
2.091	-0.04475233170953\\
2.103	-0.044058897990384\\
2.118	-0.043592991849728\\
2.168	-0.0423726897941741\\
2.183	-0.0414597503916045\\
2.197	-0.0402164394230997\\
2.21	-0.0386887547160644\\
2.223	-0.0367876698436707\\
2.236	-0.0345142966840637\\
2.249	-0.0318827809365967\\
2.263	-0.0286786438569955\\
2.278	-0.0248730600472449\\
2.296	-0.0199060171425671\\
2.323	-0.0119982257284565\\
2.351	-0.00388397349816927\\
2.368	0.000651757675980136\\
2.382	0.00401398543767861\\
2.395	0.00675874731069115\\
2.407	0.00892642932620724\\
2.419	0.0107166618333192\\
2.431	0.0121173391458296\\
2.443	0.013129301282202\\
2.455	0.0137655455904042\\
2.468	0.0140587522802633\\
2.481	0.0139817951624899\\
2.496	0.0135025047904307\\
2.513	0.0125675963698999\\
2.536	0.0108978064894885\\
2.578	0.00780258818925184\\
2.596	0.00687912591432926\\
2.612	0.00641687886697895\\
2.627	0.00634815304366221\\
2.642	0.00666428908177785\\
2.657	0.00737642588014964\\
2.672	0.00847623756168758\\
2.688	0.0100465187375622\\
2.705	0.0121062259646152\\
2.725	0.0149324185619157\\
2.753	0.019314517318243\\
2.791	0.0252178514232959\\
2.811	0.0279249238747354\\
2.828	0.0298547926508146\\
2.844	0.0312918915775633\\
2.859	0.0322700710826531\\
2.874	0.0328767900096008\\
2.889	0.0331161021020367\\
2.905	0.0329906355173133\\
2.922	0.0324768006995626\\
2.941	0.0315291371568556\\
2.966	0.0298876765016391\\
3.024	0.0259429319530238\\
3.045	0.024953964429077\\
3.065	0.0243875922422667\\
3.084	0.0242279703669563\\
3.103	0.0244418798612127\\
3.123	0.0250432256218218\\
3.145	0.0260795986725402\\
3.173	0.027788567135012\\
3.238	0.0318989240888836\\
3.261	0.0329039335001422\\
3.282	0.0334583100523318\\
3.302	0.033624720109481\\
3.322	0.0334325131862219\\
3.343	0.0328699664911696\\
3.367	0.0318470096831724\\
3.398	0.0301199801025671\\
3.467	0.0261387918143026\\
3.494	0.0250366256723584\\
3.519	0.0243916748384923\\
3.544	0.0241235915220024\\
3.571	0.02421650527282\\
3.604	0.0247272229950557\\
3.695	0.026366993754916\\
3.724	0.0263811939871186\\
3.752	0.0260206585031462\\
3.781	0.0252710251840114\\
3.815	0.0240040852036536\\
3.937	0.0190789728357039\\
3.969	0.0183882598326601\\
4.003	0.018037363690711\\
4.043	0.0180172526276365\\
4.156	0.0181909129620497\\
4.194	0.0177293673775356\\
4.235	0.0168447074515168\\
4.293	0.0151692709168234\\
4.36	0.0133158683220564\\
4.403	0.0125057250816969\\
4.446	0.0120708068198336\\
4.5	0.011925763215439\\
4.603	0.0117161550457183\\
4.655	0.0111656769249908\\
4.718	0.0100952566445294\\
4.818	0.00838957886218594\\
4.873	0.00787051481293233\\
4.937	0.00766594535471832\\
5.071	0.0073414478099707\\
5.145	0.00667242223184772\\
5.284	0.00533280581389306\\
5.358	0.00508701616313267\\
5.563	0.00465073723066567\\
5.771	0.00364251845036989\\
6.126	0.00296833356675741\\
6.252	0.0027999535174601\\
6.487	0.00258157844392315\\
6.695	0.00230987993520415\\
6.982	0.00207446173863879\\
7.236	0.00192474236402518\\
8.001	0.00140037095395584\\
};
\addlegendentry{$Q^g_{G\text{,RFC1}}$}

\addplot [color=red, line width=1.1pt]
  table[row sep=crcr]{%
1.499	0\\
1.8	0\\
1.801	0.862543960090314\\
1.807	0.854906446002337\\
1.813	0.847831950747073\\
1.819	0.841268928990056\\
1.826	0.834197313753723\\
1.833	0.827694453893878\\
1.84	0.821703309895762\\
1.847	0.816173696400694\\
1.855	0.810362827576823\\
1.863	0.805039552110314\\
1.871	0.800153885413971\\
1.88	0.795126811384504\\
1.889	0.790545629279322\\
1.898	0.786364839665572\\
1.908	0.782141919529149\\
1.918	0.778319856120687\\
1.929	0.774532870552557\\
1.94	0.771141485985282\\
1.952	0.767850841812313\\
1.964	0.76494931703755\\
1.977	0.762207717314523\\
1.99	0.759850763552635\\
2.001	0.758135719641613\\
2.002	0.0817466077628151\\
2.008	0.0768307791011154\\
2.014	0.072393919703682\\
2.02	0.0684082259064631\\
2.026	0.064845975668284\\
2.032	0.0616796901551382\\
2.038	0.0588822511944063\\
2.045	0.0560491905440834\\
2.052	0.0536403722716976\\
2.059	0.0516152672163859\\
2.067	0.0497200751150508\\
2.075	0.048217171966316\\
2.084	0.0469262998126005\\
2.094	0.0458995684462558\\
2.105	0.0451516642467951\\
2.119	0.0445918293263663\\
2.148	0.0439431342014664\\
2.167	0.0432907911431641\\
2.182	0.0423973616743627\\
2.195	0.0412660160663698\\
2.208	0.0397604058191412\\
2.221	0.0378612575212465\\
2.234	0.0355677192726969\\
2.247	0.0328954451561447\\
2.261	0.0296293247166837\\
2.276	0.0257451252800234\\
2.294	0.0206818626949516\\
2.323	0.0120608283263888\\
2.348	0.00476641264785016\\
2.365	0.000193488405768605\\
2.379	-0.00320202845394668\\
2.392	-0.00598598197268174\\
2.404	-0.00820100483415054\\
2.416	-0.0100518291611138\\
2.428	-0.011527258346451\\
2.44	-0.0126273169264213\\
2.452	-0.013362488617398\\
2.465	-0.013770436045629\\
2.478	-0.013809739722717\\
2.492	-0.0134922667712445\\
2.508	-0.0127569294081109\\
2.528	-0.0114445161374288\\
2.59	-0.00706460380449236\\
2.607	-0.00640101199636156\\
2.622	-0.00617395412916366\\
2.637	-0.00632867800025316\\
2.651	-0.00683704623172687\\
2.665	-0.00769755185706522\\
2.68	-0.00899216561326632\\
2.696	-0.0107555897824643\\
2.714	-0.0131294964074726\\
2.736	-0.0164229556684354\\
2.798	-0.0259430168210848\\
2.816	-0.0282187626818047\\
2.833	-0.0299897801138656\\
2.849	-0.0312684717828109\\
2.864	-0.0321002388917009\\
2.879	-0.0325706650997688\\
2.895	-0.0326859781415685\\
2.911	-0.0324332849737399\\
2.929	-0.0317700899884432\\
2.95	-0.03060641903887\\
2.981	-0.0284652640903857\\
3.02	-0.02583640517493\\
3.042	-0.0247274901775807\\
3.062	-0.0240901221549983\\
3.081	-0.0238610071147214\\
3.1	-0.0240104232790301\\
3.12	-0.0245543586796728\\
3.142	-0.0255430106713739\\
3.169	-0.0271501936519272\\
3.24	-0.0315849254453973\\
3.263	-0.0325392994127132\\
3.284	-0.0330466666648483\\
3.304	-0.0331736747965135\\
3.325	-0.0329314427960483\\
3.347	-0.032297243890854\\
3.372	-0.0311889200327915\\
3.406	-0.0292644332931928\\
3.464	-0.0259513293314715\\
3.491	-0.0248141704190719\\
3.516	-0.0241294773750731\\
3.541	-0.0238228785605994\\
3.568	-0.0238822982276243\\
3.6	-0.0243496300333987\\
3.699	-0.0260937268842252\\
3.728	-0.0260583825653224\\
3.756	-0.0256537590429566\\
3.786	-0.0248383810813966\\
3.822	-0.0234621834189603\\
3.927	-0.0191792491193823\\
3.96	-0.0183668268395696\\
3.993	-0.0179325629775118\\
4.03	-0.0178271064885855\\
4.098	-0.0181068617285369\\
4.147	-0.0180990757143586\\
4.186	-0.0177159479300997\\
4.226	-0.0169392851600403\\
4.277	-0.0155430123200642\\
4.363	-0.0131613640631869\\
4.406	-0.0123852362945556\\
4.45	-0.011975354389433\\
4.507	-0.0118575317984497\\
4.6	-0.0116747362214742\\
4.652	-0.011151109669358\\
4.714	-0.0101239826407529\\
4.82	-0.00833111997622282\\
4.875	-0.00782919105947144\\
4.94	-0.0076359918440474\\
5.069	-0.00733015911069579\\
5.142	-0.00668501937371069\\
5.287	-0.00530292068092209\\
5.362	-0.00507083337323166\\
5.556	-0.00468000716884021\\
5.782	-0.00362336346424286\\
6.071	-0.00314460176824838\\
6.214	-0.0028117193482462\\
6.963	-0.00209829417896223\\
7.215	-0.00193192491126304\\
7.818	-0.00151937184117479\\
8.001	-0.00139944909013501\\
};
\addlegendentry{$Q^g_{G\text{,RFC2}}$}

\end{axis}
\end{tikzpicture}%
		\caption{Zoom of reactive power.} \label{Fig 16: 1 phase Re Power zoom}
	\end{subfigure}
	\caption[]
	{\small Case 2 results of generated active and reactive power of each RFC's generator.} \label{Fig 13: Case 2 part 2}
\end{figure*}

The average frequency of the relative rotor oscillation, $\Delta\omega_{12}=\omega_{\text{RFC1}}-\omega_{\text{RFC2}}$, between the RFC in interconnected mode is about 2.27~Hz see \Cref{Fig 10: Case 2 relative rotor speeds}. The single-phase power between the RFCs oscillates with same frequency, see \Cref{Fig 13: 1 phase Ac Power,Fig 14: 1 phase Ac Power zoom}.

Note that the average frequency of the single-phase active power oscillations differs from the three-phase active power oscillation. There are several reasons for the difference, such as the grid impedance of the railway grid, rotor angle difference between RFC single-phase generators, location of the fault and excitation system equipped on the single-phase generators.

%The frequency of the single-phase active power oscillations are reasonable, as they occur when synchronous machines swing against each other inside a power station or with nearby power station according \cite{Basler2005a,Rogers2000}.

As in Case 1, the three-phase power consumed by each individual RFC motor lags the single-phase active power generated by the RFC generator, see \Cref{Fig 12: Case 2 part 1}.

The simulation shows also that the reactive power oscillates with a frequency of approximately 2.45~Hz in the single-phase side, see \Cref{Fig 15: 1 phase Re Power,Fig 16: 1 phase Re Power zoom}.

\clearpage

\section{Conclusions} \label{Sec: Conclusion}

An open high order electromechanical dynamical model of an RFC has been developed by using well established synchronous machine models. The model is suited for electromechanical stability studies of low frequency railway grids that are synchronously connected to three-phase public grids.

Numerical studies in the phasor domain have been performed, to graphically explain and illustrate the behaviour of the proposed RFC model, using data for one of the most common RFC models (the Q48/Q49) used in Sweden. The case studies include both single feeding of a catenary section and two-sided catenary feeding (that is, the simples way that a railway power system can be operating in interconnected mode). The two basic cases of feeding a railway together facilitates illustrating many important aspects of the RFC behaviour in simple manners. From the modelling and the case studies done, it can be concluded that the three-phase active power lag the single-phase active power, in line with \cite{Danielsen2010}, as the inertia of the RFC will cause a time delay.

%From the Case study 1, when the RFC is operating in single feeding mode, it was found that the dominant frequency of rotor speed oscillation is 1.96~Hz for the RFC type used. The three-phase active power from the RFC motor oscillates with the same frequency, and the system is shown to be stable as rotor speed oscillation approach zero after a few cycles.
%
%In Case study 2, when the RFC are operating in interconnected mode, both active and reactive power oscillations occur between the RFCs. The active power supplied by each RFC generator oscillates in opposite phase with each other. The three-phase power does not oscillate with the same frequency as the single-phase power, as grid impedance, excitation system and relative angles between the RFCs generators affect the single-phase active and reactive power oscillations.

%The rotor oscillations occurs with the RFC model proposed in both cases are reasonable, as it falls in the 0.1-3~Hz range of low-frequency rotor oscillations \cite{Kundur1994,Basler2005a,Li2016book,RobustControll2005}.

The model presented could be used for electro-magnetic transient (EMT) simulations with proper adaptations, depending on the desired level of details. However, going into deeper details may be a complicated task, specially how to adapt the electrical machine equations for the single-phase salient pole generator. The reader should note, that single-phase powers (active as well as reactive), unlike three-phase powers, are not constant even in steady-state operation. Single-phase power oscillate with twice the grid frequency.

The open models presented in this work constitute an important step in the work for ensuring the future reliability and stability of low-frequency railways, an important societal asset. The high order RFC model in this paper is the first of its kind presented in the academic literature. Its implementation for stability studies done in this paper provide an open starting platform, which can be used for education and future research of low-frequency railway grids synchronously connected to three-phase public grids. 

\section*{Acknowledgement}
The financial support for this project from the Swedish Transport Administration is greatly acknowledged.
%\newpage

%\input{Sections/ConclusionsOLD}

%\newpage

%\clearpage



%% References


\bibliographystyle{model3-num-names}
\bibliography{library}
% Encoding: UTF-8

%@Comment{jabref-meta: databaseType:bibtex;}

%% not want to use model3-num-names.bst.





% \end{thebibliography}

%\end{linenumbers}

\end{document}